\documentclass[twocolumn]{aastex63}

\turnoffedit

\graphicspath{{./}{Figures/}{Figures/Geometry/}}

\usepackage{booktabs}

\shorttitle{Properties of streamer wave events}
\shortauthors{B. Decraemer et al.}

\begin{document}

\title{Properties of Streamer Wave Events Observed During the \textit{STEREO} Era}

\correspondingauthor{Bieke Decraemer}
\email{bieke.decraemer@oma.be}

\author[0000-0002-1335-7768]{Bieke Decraemer}
\affiliation{Solar-Terrestrial Centre of Excellence - SIDC, Royal Observatory of Belgium, Ringlaan 3, 1180 Brussels, Belgium}
\affiliation{Centre for mathematical Plasma Astrophysics, Mathematics Department, KU Leuven, Celestijnenlaan 200B, 3001 Leuven, Belgium}

\author[0000-0002-2542-9810]{Andrei N. Zhukov}
\affiliation{Solar-Terrestrial Centre of Excellence - SIDC, Royal Observatory of Belgium, Ringlaan 3, 1180 Brussels, Belgium}
\affiliation{Skobeltsyn Institute of Nuclear Physics, Moscow State University, 119992 Moscow, Russia}

\author[0000-0001-9628-4113]{Tom Van Doorsselaere}
\affiliation{Centre for mathematical Plasma Astrophysics, Mathematics Department, KU Leuven, Celestijnenlaan 200B, 3001 Leuven, Belgium}

\begin{abstract}
Transverse waves are sometimes observed in solar helmet streamers, typically after the passage of a coronal mass ejection (CME). The CME-driven shock wave moves the streamer sideways, and a decaying oscillation of the streamer is observed after the CME passage. Previous works generally reported observations of streamer oscillations taken from a single vantage point (typically the \textit{SOHO} spacecraft). We conduct a data survey searching for streamer wave events observed by the COR2 coronagraphs onboard the \textit{STEREO} spacecraft. For the first time, we report observations of streamer wave events from multiple vantage points, by using the COR2 instrument on both \textit{STEREO A} and \textit{B}, as well as the \textit{SOHO}/LASCO \edit1{\added{C2+C3}} coronagraph\edit1{\added{s}}. We investigate the properties of streamer waves by comparing the different events and performing a statistical analysis. Common observational features give us additional insight on the physical nature of streamer wave events. The most important conclusion is that there appears to be no relation between the speed of the CME and the phase speed of the resulting streamer wave, indicating that the streamer wave speed is determined by the physical properties of the streamer rather than the properties of the CME. This result makes streamer waves events excellent candidates for coronal seismology studies. From a comparison between the measured phase speeds and the phase speeds calculated from the measured periods and wavelengths, we could determine that the speed of the post-shock solar wind flow in our streamers is around 300 $\mathrm{km \ s}^{-1}$.
\end{abstract}




\section{Introduction}\label{s:intro}
Coronal streamers are quasi-static ray-like structures extending from the lower to the outer solar corona and appearing most strikingly during total eclipses \citep[e.g. ][]{loucif_solar_1989, koutchmy_1992}. Streamers can persist for months and extend up to 30 solar radii as seen in coronagraphic images. In white-light images, the narrowly outlined bright streamer stalk is actually the head-on projection of a dense plasma sheet \citep[see e.g. ][]{wang_1997, saez_2007, decraemer_2019}. This plasma sheet envelops a current sheet, that extends outwards as the heliospheric current sheet. Helmet streamers are thus typically found above active regions and filament channels \citep{newkirk_structure_1967, koutchmy_three_1971, zhukov_origin_2008}. The bright and dense structure of the streamer traces out the coronal magnetic field and can give insight to the global magnetic configuration of the corona \citep{pneuman_1971, wang_2000, saez_2005, saez_2007, lamy_2019}.

The quasi-static streamers are perturbed by dynamic events such as coronal mass ejections \citep[CMEs, see e.g. ][]{webb_2012}. These are violent energetic eruptions in the solar atmosphere, consisting of large outward-propagating structures containing plasma and magnetic fields. They can cause large-scale disturbances in the corona on a timescale from minutes to hours. CMEs are a fundamental mechanism to remove the built-up magnetic energy and plasma from the large-scale solar corona. The understanding of the mechanism behind CMEs has challenged solar physicists for over four decades. A rapid release of magnetic energy is the only energy source that can lead to the explosive properties of CMEs. Different models exist on how this magnetic energy is released \citep{forbes_2000}, but none can yet be proclaimed to be the correct and complete model.

Since coronal streamers and CMEs are two very commonly observed features in the solar corona, one can expect that they interact on a regular basis. Most often, CMEs build up and erupt from inside a pre-existing coronal streamer \citep[see e.g. ][]{hundhausen_1993}. The streamer is then, at least partially, disrupted. Of more interest to this paper is another kind of interaction, when the streamer is affected by a lateral expansion of a CME originating outside of it and gets hit by either the expanding CME itself or by associated disturbances like a CME-driven shock wave. This interaction can deflect the streamer stalk and cause transverse disturbances in the streamers. The disturbances can be used for the tracing of the CME-driven shock \citep{sheeley_shock_2000}. When a streamer is hit sideways by a CME, magnetic reconnection along the current sheet may also be triggered. This has been observed by indications of streamer detachment \citep{koutchmy_1973, sheeley_inout_2007, bemporad_2008, bemporad_2010}, the release of plasma blobs \citep{song_2012}, and the formation of streamer in/out pairs \citep{sheeley_inout_2007}. Several studies combining white-light and radio observations have also revealed that CME-streamer interactions are an important source of type II radio bursts \citep{reiner_2003, cho_2007, cho_2008, feng_2012, kong_2012, magdalenic_2014}. A more recent discovery in white-light observations are the so-called streamer waves \citep{chen_streamer_2010, feng_streamer_2011, feng_2013, kwon_2013}. They are described as outward-propagating wavy motions of the streamer stalk, excited by the interaction of a rapidly moving and expanding CME with a nearby coronal streamer. They are one of the largest wave phenomena ever observed in the solar corona and the even largest resolved periodic waves in the solar system. 

Magnetohydrodynamic (MHD) waves and oscillations have already been a topic of study in the solar corona for many decades. The very dynamic solar atmosphere is an ideal environment for the generation and propagation of waves on all relevant temporal and spatial scales. Wave phenomena are produced by perturbations of the plasma parameters and magnetic field. Waves are often thought of as a possible explanation for coronal plasma heating \citep[see e.g. ][and references therein]{arregui_2015}, solar wind acceleration \citep{ofman_2010, cranmer_2012}, and as a possible mechanism for quasi-periodic pulsations in solar flares \citep{dolla_2012, vandoorsselaere_2016, mclaughlin_2018}. Combining observations of waves with the numerous theoretical studies resulted in a very powerful technique for diagnosing the plasma parameters of the coronal medium through which the waves propagate. This technique is now referred to as coronal seismology \citep[see e.g. reviews by][and references therein]{nakariakov_2005, andries_2009, demoortel_2012}. \edit1{\added{One very well-known example are transverse loop oscillations. Observations of these transverse waves \citep[see e.g. ][]{nakariakov_1999,white_2012}, prove that they are a common occurrence in eruptive events in the solar corona. The physics of these transverse waves is probably similar to that of streamer waves, except for the presence of the current sheet. By measuring the period and the wavelength of a transverse loop oscillation, \citet{nakariakov_2001} demonstrated that they could estimate the local magnetic field strength. As some physical parameters, such as the magnetic field strength, are difficult to measure directly in the solar corona, coronal seismology has become a popular method to obtain reliable values for these parameters with improving models and observations \citep[see e.g.][and references therein]{west_2011, wang_2015, prasad_2018, pascoe_2019}.}} 

Previous works reported observations of streamer oscillations taken from a single vantage point \citep[typically the \textit{SOHO} spacecraft; ][]{chen_streamer_2010, feng_streamer_2011, feng_2013} or only mentioned the streamer oscillation as a secondary event \citep{kwon_2013}. Except for \citet{feng_2013}, these works report the streamer wave as a decaying oscillation of the streamer after the CME-driven shock wave moves the streamer sideways. The magnetic field of the streamer provides the restoring force to support the wave-like motion that is observed in the streamer stalk, after the deflection of the streamer by the CME impact. The streamer wave observed by \citet{feng_2013} was not due to an impulsive excitation (there is no observational indication of a CME near the streamer), but was probably caused by the Kelvin–Helmholtz instability. Streamer waves are usually interpreted as a fast body kink MHD mode, which propagates outward along the plasma sheet of the thin streamer stalk. All the observed streamer waves decay in just a few periods.
  
In this paper, we present a data survey searching for the streamer wave events observed by the COR2 coronagraphs aboard the twin spacecraft of the \textit{Solar Terrestrial Relations Observatory} (\textit{STEREO}) mission \citep{kaiser_stereo_2008}. Each coronagraph is a part of the Sun Earth Connection Coronal and Heliospheric Investigation (SECCHI) instrument package \citep{howard_sun_2008}. The \textit{STEREO}/COR2 data are complemented by the data taken by the Large-Angle Spectroscopic Coronagraph \citep[LASCO, see ][]{brueckner_large_1995} \edit1{\added{C2 and C3 telescopes}} aboard the \textit{Solar and Heliospheric Observatory} (\textit{SOHO}). In Section~\ref{s:survey}, we describe the selection method for the data survey and discuss the observalibility of streamer wave events. We describe how we performed the measurements of streamer wave properties in Section~\ref{s:measure}. In Section~\ref{s:stats}, a statistical analysis of the measurements for our data set is carried out. A discussion and our conclusions are presented in Section~\ref{s:concl}.

\section{Data survey}\label{s:survey}

\begin{figure*}
\centering
\includegraphics[width=0.95\textwidth]{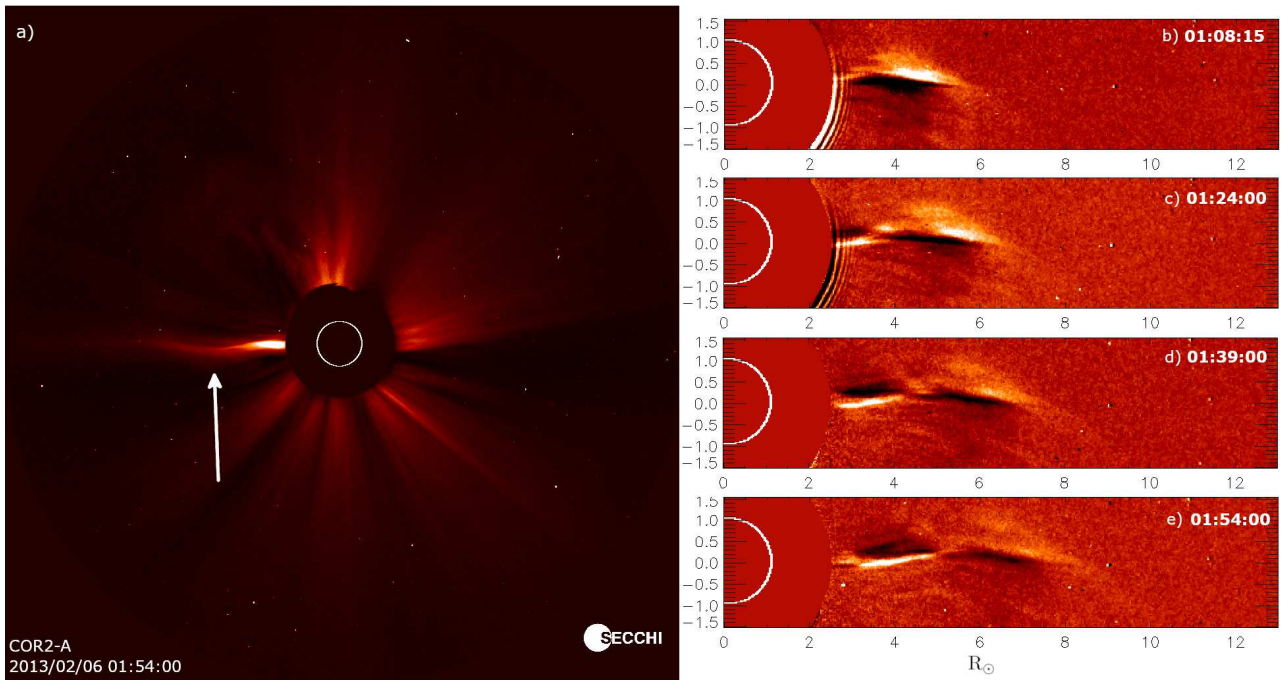}
\caption{Streamer wave event 13 observed on 2013-02-06 with \textit{STEREO} A/COR2. Panel (a) shows a screenshot of the full field-of-view of COR2 A in white light during the streamer wave event, where the streamer wave is indicated by the arrow. Panels (b) to (e) present running difference images of the streamer rotated clockwise by 180\degr{} at 4 times during the event.}
\label{fig:event20130206}
\end{figure*}

We first examined the \textit{STEREO A} and \textit{B} COR2 data for streamer oscillation events from the start of the \textit{STEREO} mission in January 2007 until the end of September 2017. Then, we checked if the streamer wave is also observed in the data taken by LASCO \edit1{\added{C2 and C3}} aboard \textit{SOHO}. In this section we present the method we used to find all events in this time frame and the 22 events that we found.

We first made a selection of candidate streamer oscillation events by going through the white-light observations of the COR2 coronagraphs aboard \textit{STEREO A} and \textit{B}. In the white-light images, we visually identified a candidate event when a spatial signature of a wavy motion was present in a coronal streamer, as can be seen in Figure \ref{fig:event20130206}(a). We did not restrict our survey to examining only time frames around CMEs reported in a given catalog, as in \citet{feng_streamer_2011}. This way, we wanted to prevent missing any oscillatory events that could arise due to other factors than the CME-streamer interaction. The process gave us a total of 68 candidate events. We narrowed down this list by carefully examining the running difference images for each candidate event. A candidate event is selected as a real event when there is a clear oscillatory signature visible in the running difference images, that is, if at least 2 alternating groups of black/white patches are visible as can be seen in Figure \ref{fig:event20130206}(c), (d), and (e). In most of the candidate events the streamer is simply displaced and there is no clear oscillatory signature. In 22 events, however, an oscillatory signature was identified and these events, presented in Table \ref{tab:events}, are investigated in our study. \edit2{\deleted{We will provide an online catalog for the events reported in this study.}}For all these events, we could identify a CME that disturbed the streamer. Therefore, we conclude that all our streamer wave events result from a coronal streamer being disturbed by a CME. It thus seems that the case reported by \citet{feng_2013} is a unique event.

Looking at the dates of the events, we notice that there are significantly more events during the solar maximum period (2011-2013) than in other years. We do have to note here that we can not find events during the period from December 2014 until November 2015, since there is no \textit{STEREO} data available for this time. After this data gap, we only have \textit{STEREO A} data available, so we also could have less events during this time because of the missing data of \textit{STEREO B}. It is however understandable that more streamer wave events occur during the solar maximum period, since there is a greater chance to observe streamers and CMEs interacting with each other during a period of high activity.

\begin{figure*}
\centering

\includegraphics[width=0.31\textwidth]{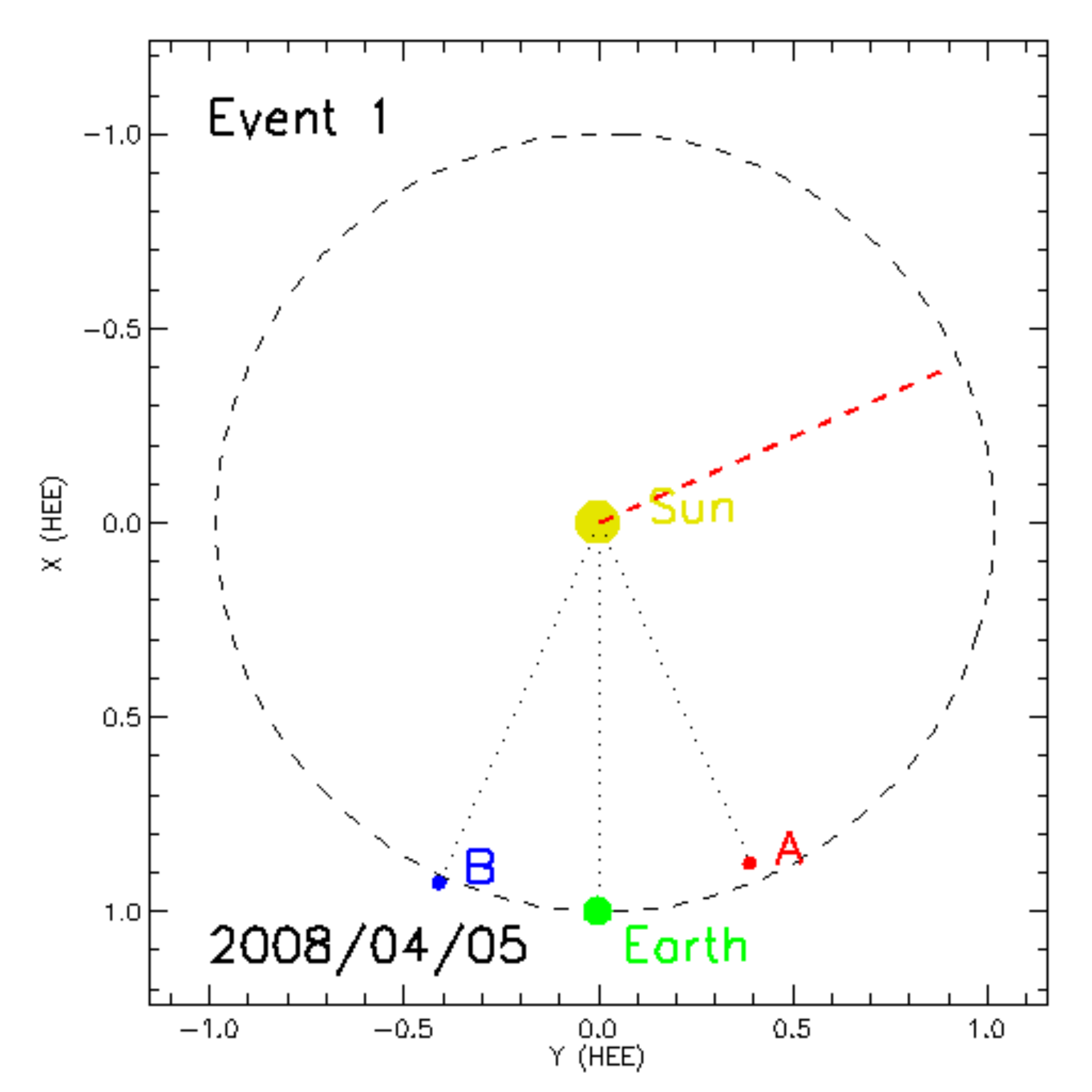}~
\includegraphics[width=0.31\textwidth]{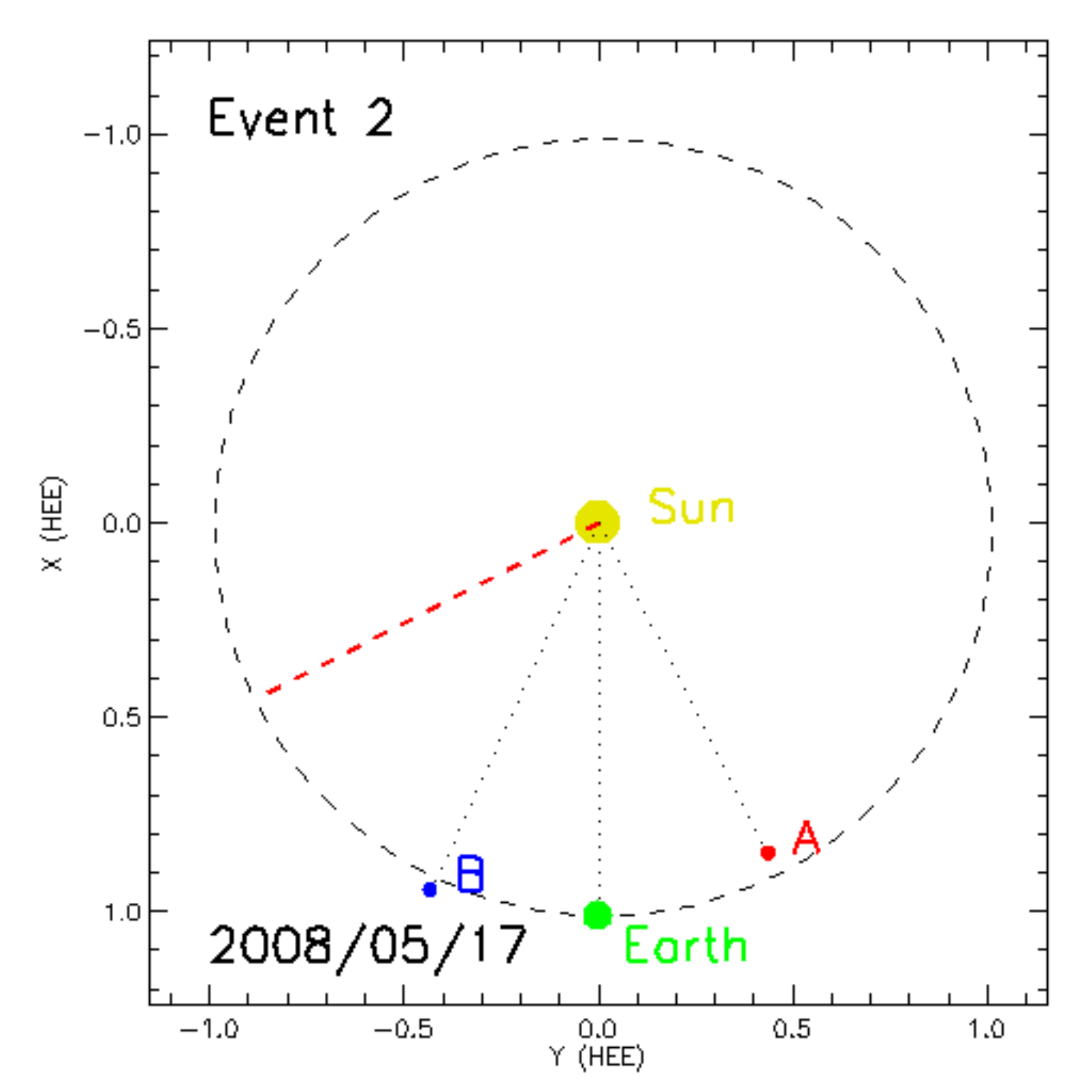}~
\includegraphics[width=0.31\textwidth]{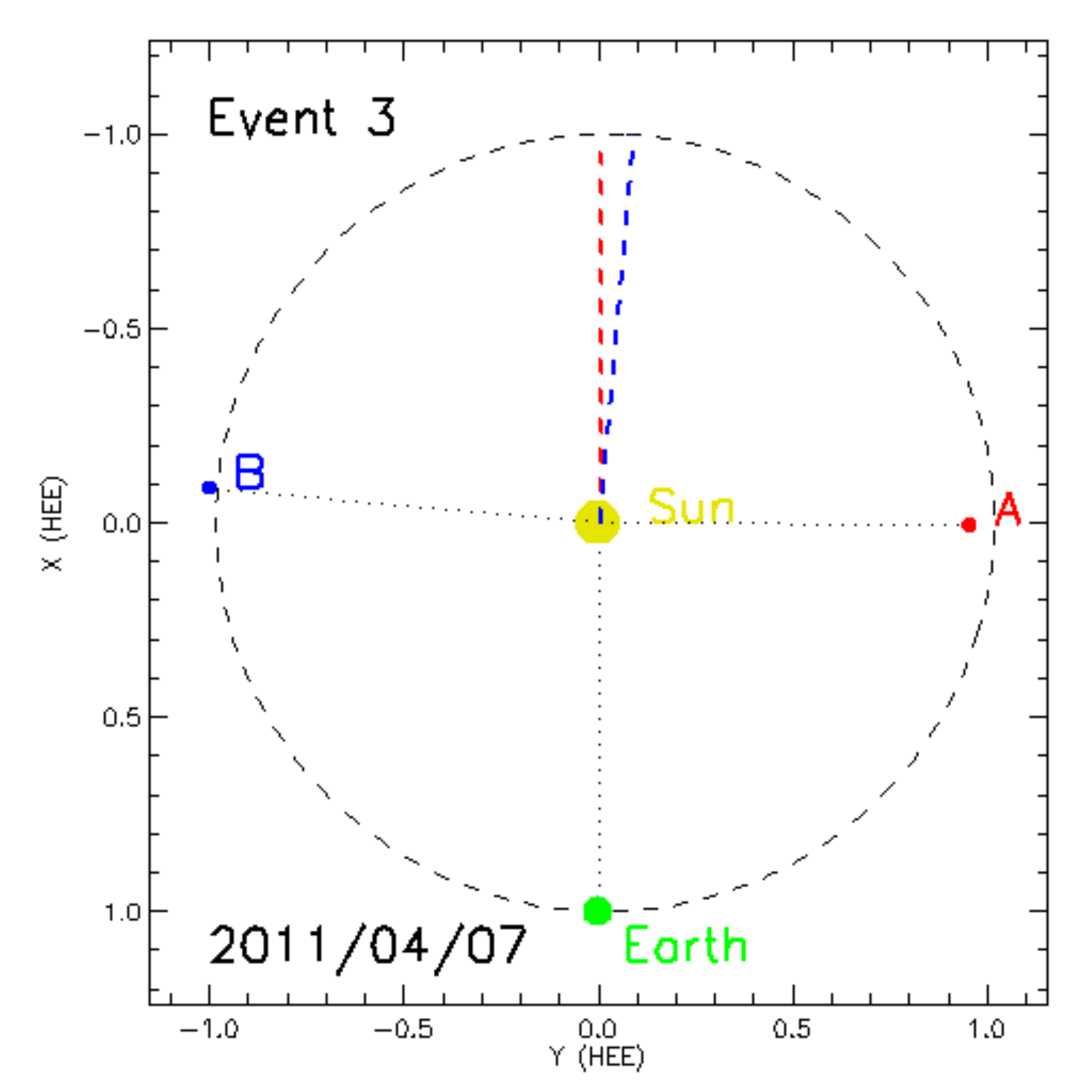}\\
\includegraphics[width=0.31\textwidth]{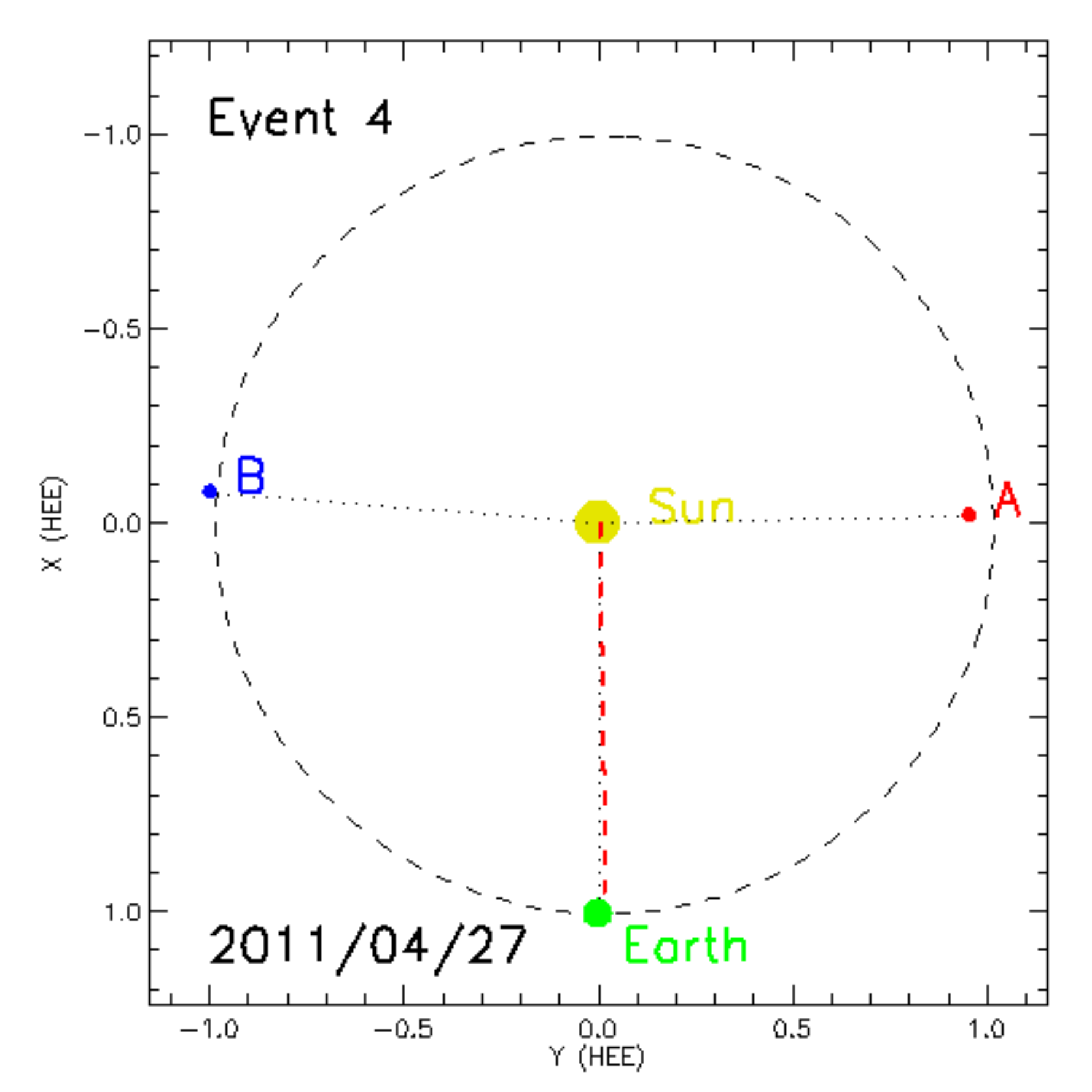}~
\includegraphics[width=0.31\textwidth]{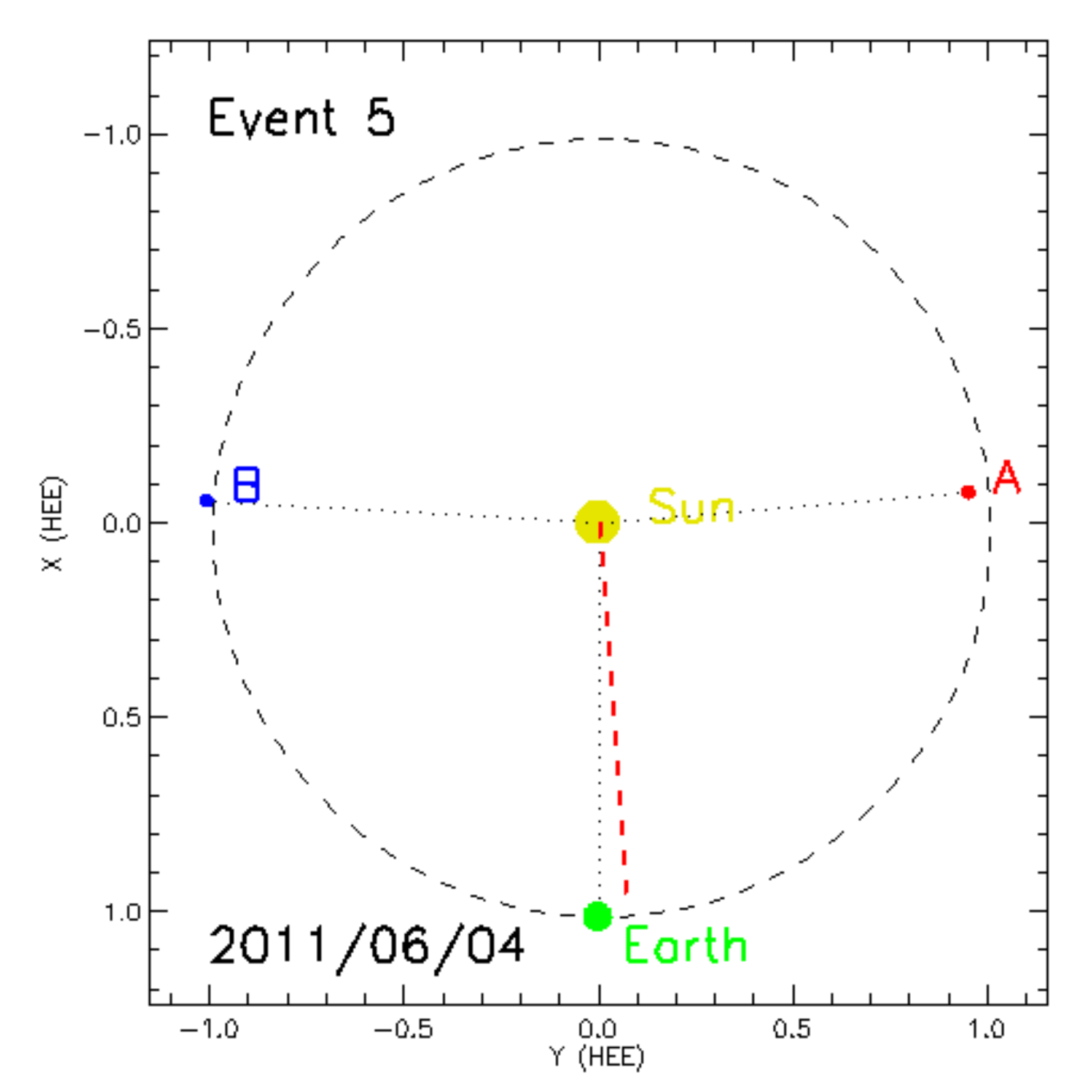}~
\includegraphics[width=0.31\textwidth]{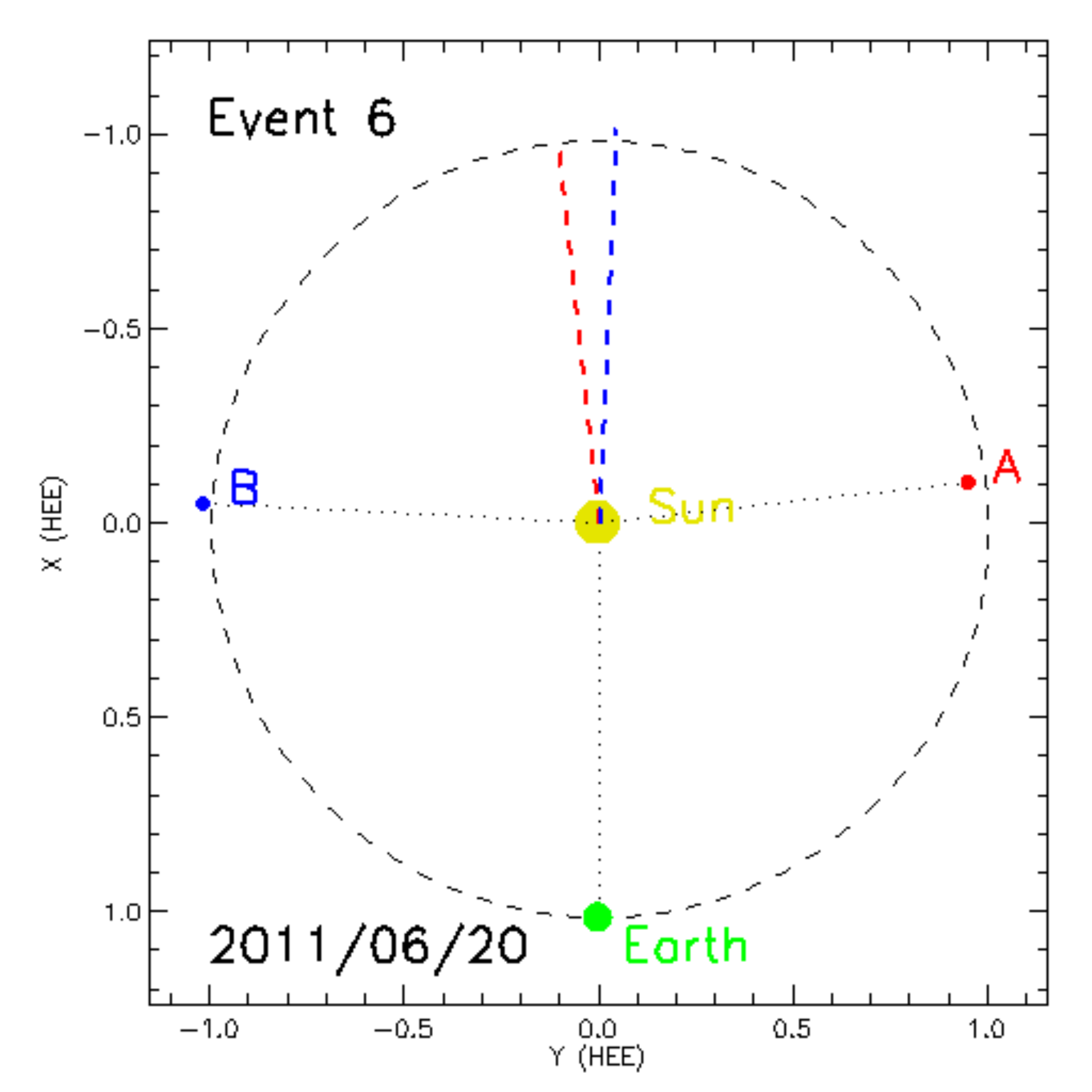}\\
\includegraphics[width=0.31\textwidth]{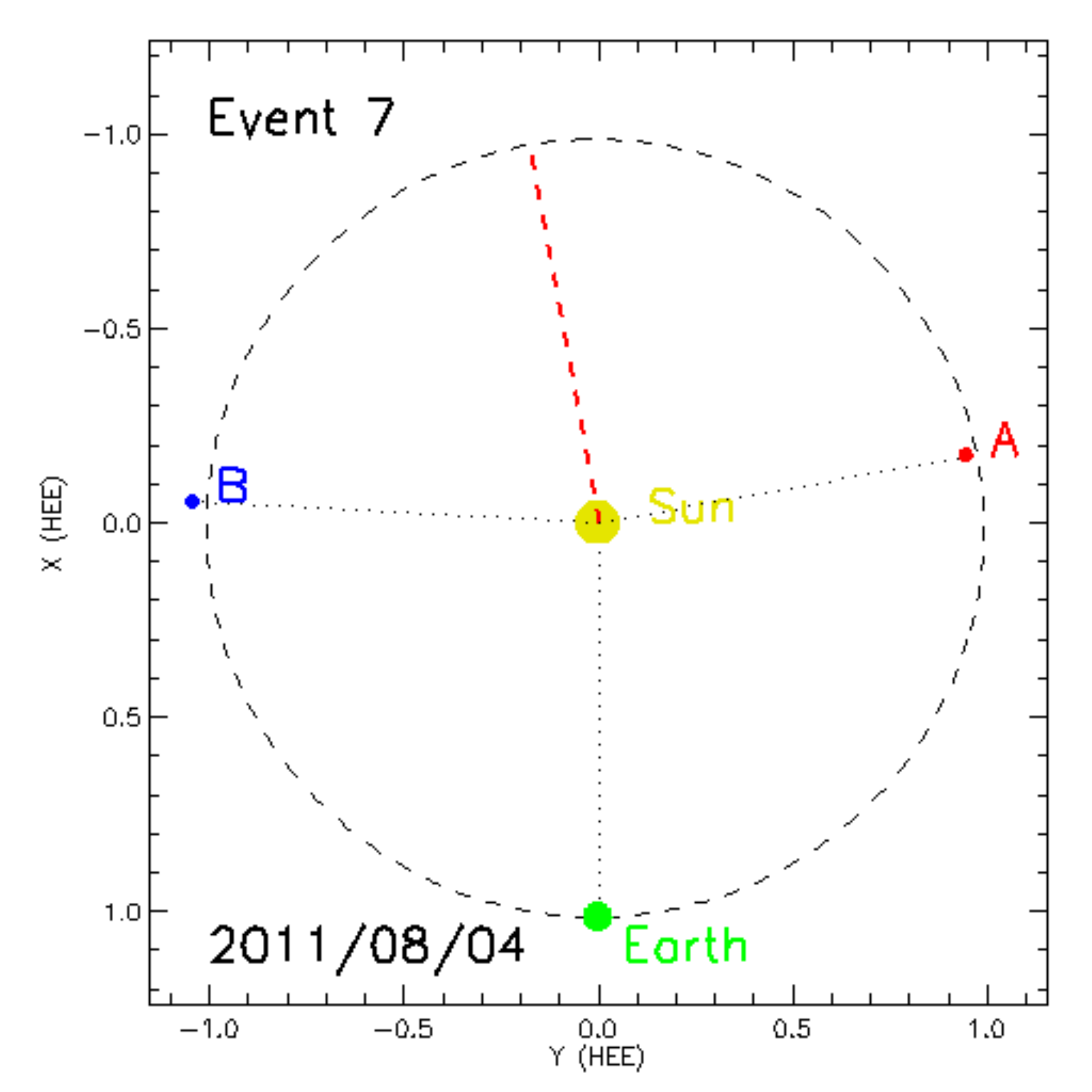}~
\includegraphics[width=0.31\textwidth]{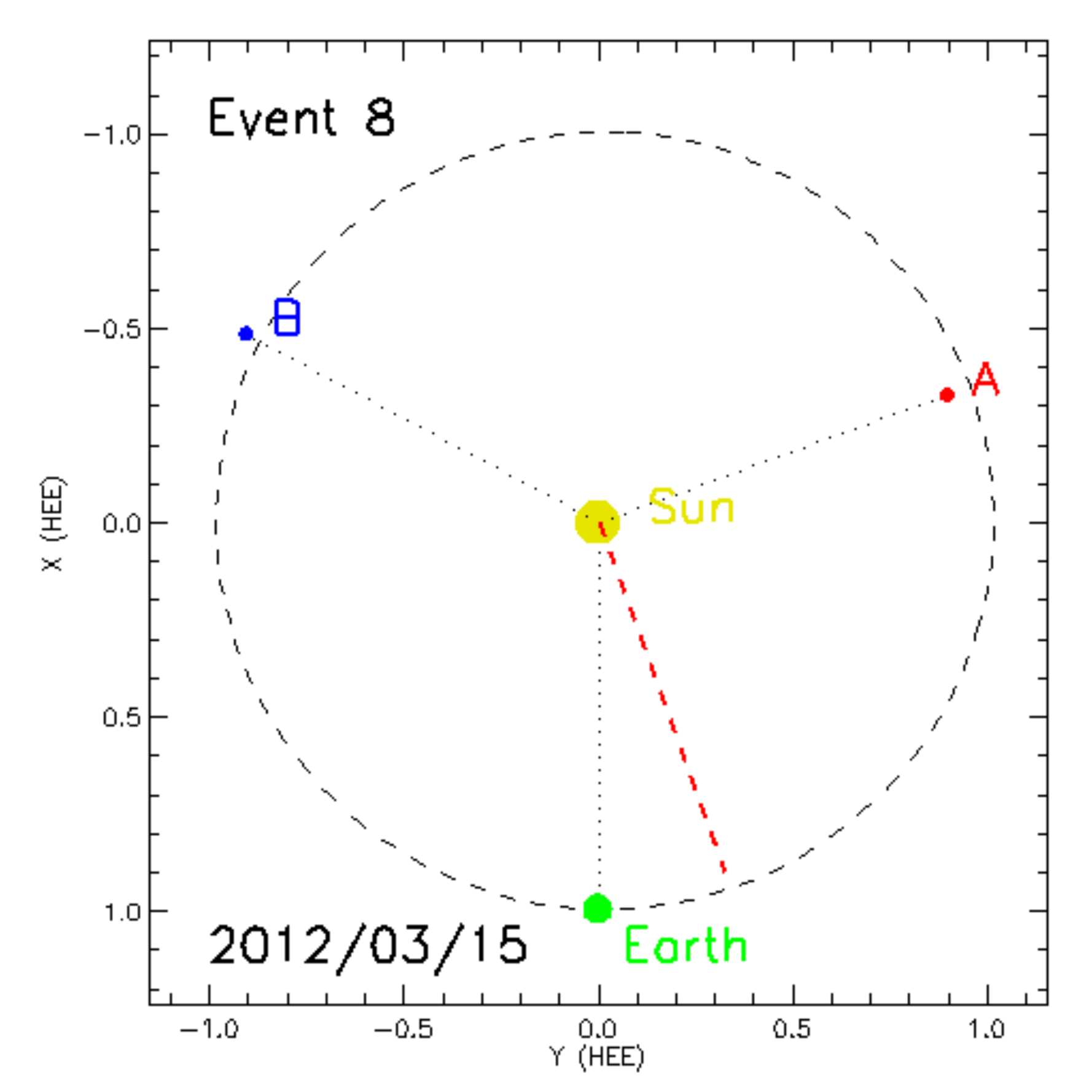}~
\includegraphics[width=0.31\textwidth]{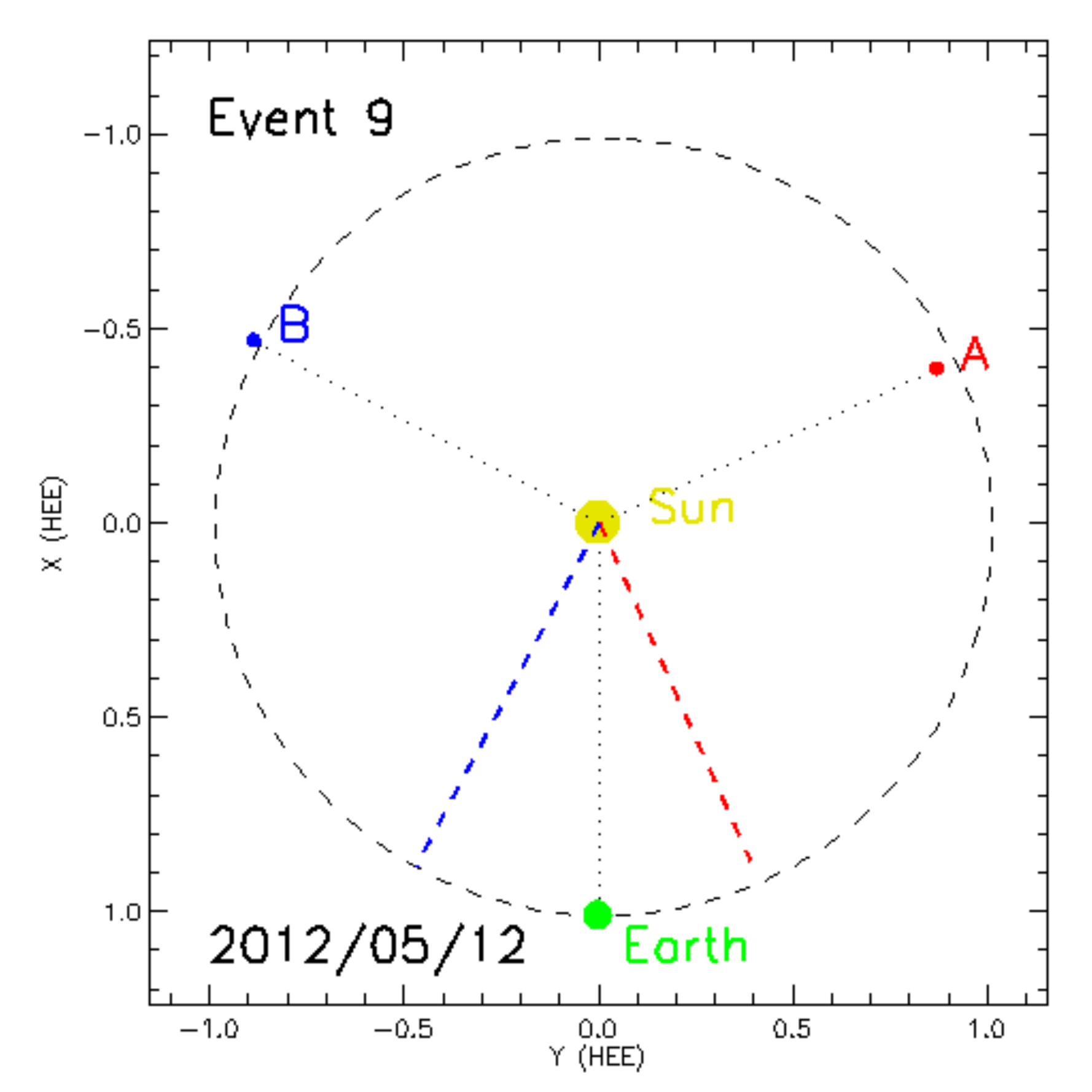}\\
\includegraphics[width=0.31\textwidth]{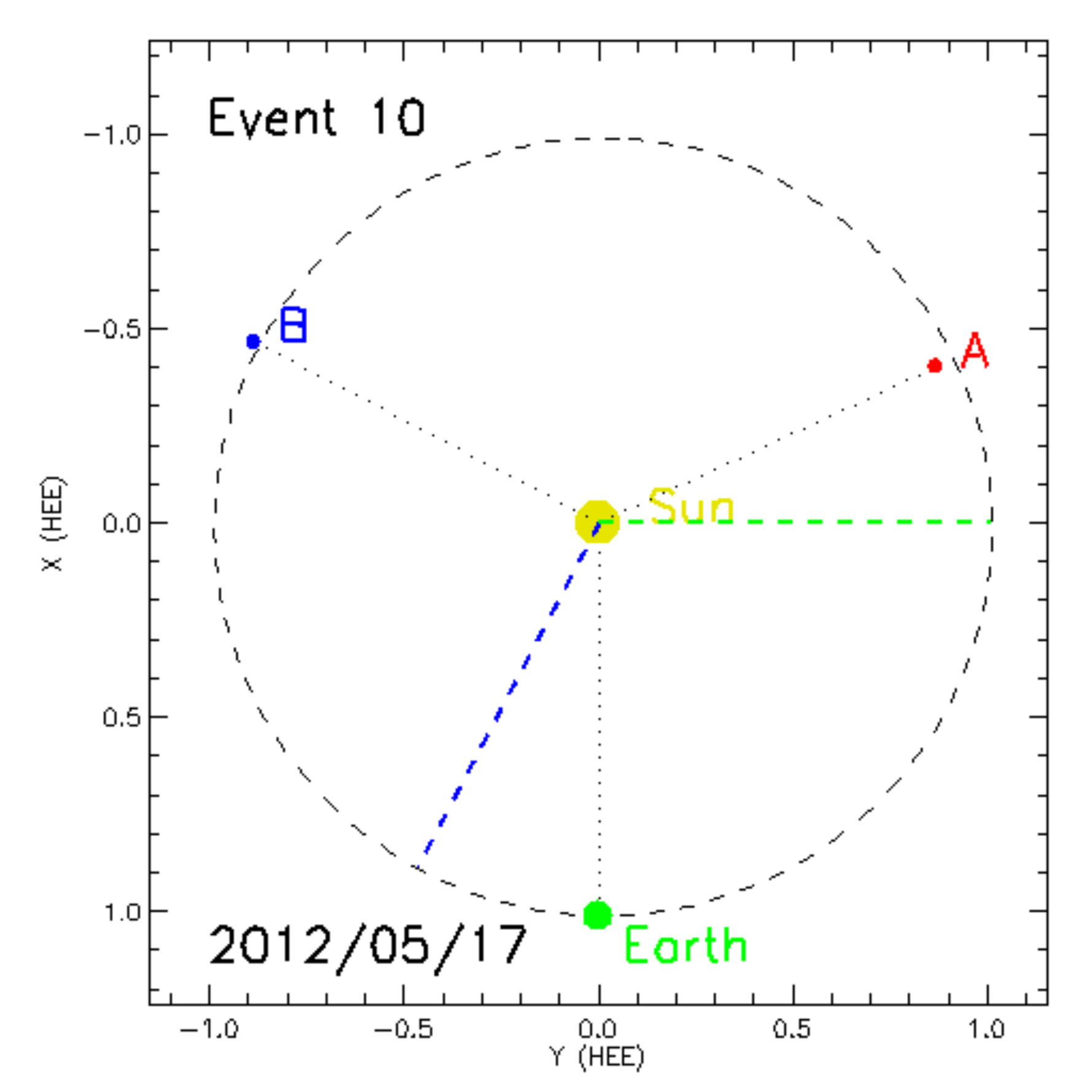}~
\includegraphics[width=0.31\textwidth]{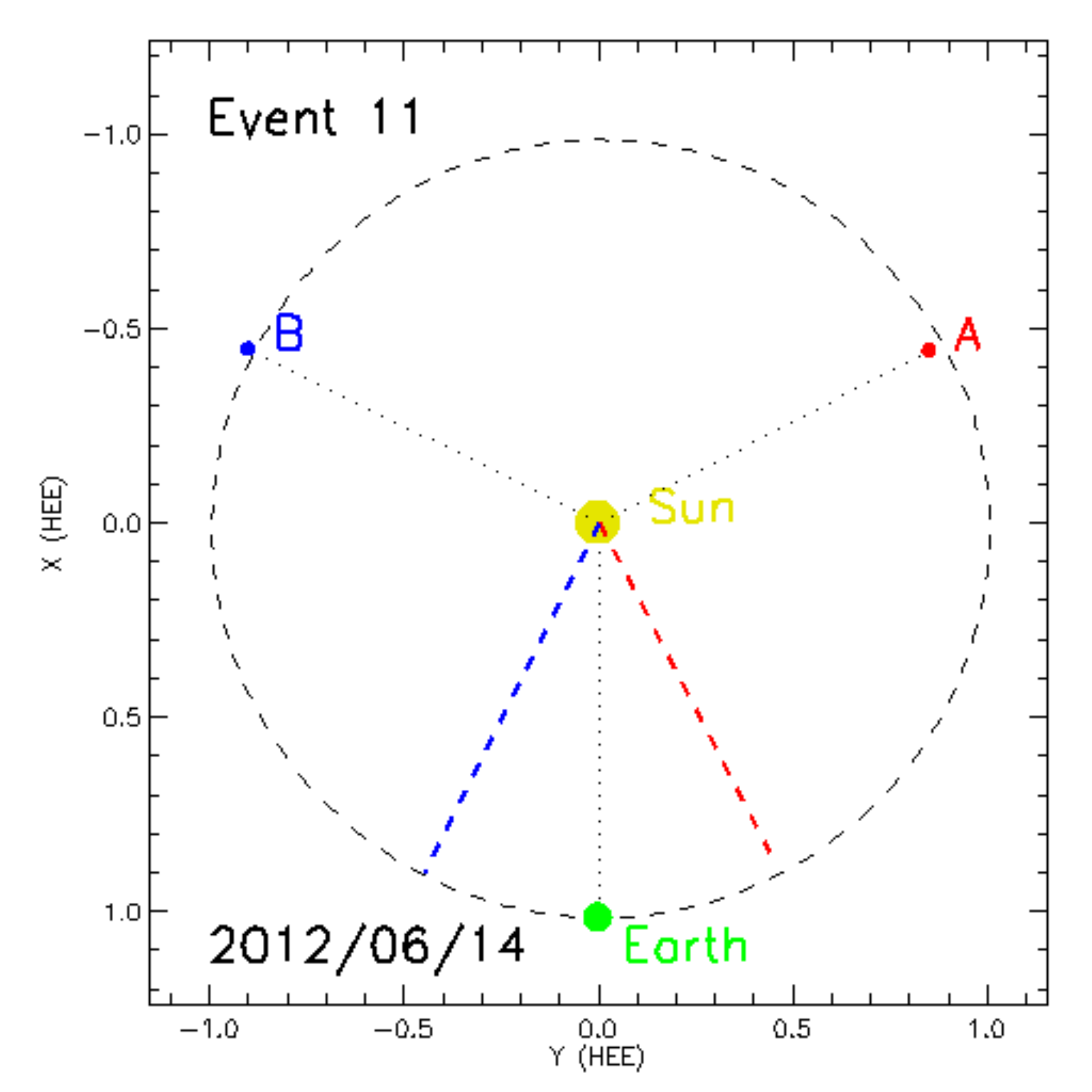}

\caption{Positions of \textit{STEREO A} and \textit{B}, and Earth (approximately the same location as \textit{SOHO}) in the heliocentric Earth ecliptic (HEE) coordinate system for all streamer wave events. If a coronagraph observed the streamer wave event, the corresponding plane of the sky above the limb on which the streamer is observed is indicated with a dashed line. The color corresponds to different spacecraft: red for \textit{STEREO A}, blue for \textit{STEREO B}, and green for Earth/\textit{SOHO}. The yellow circle is the Sun. For event 12, the streamer has a PA of 0\degr{}, which means it is observed exactly at solar north. This is indicated by a blue asterisk on top of the Sun.}
\label{fig:positions}
\end{figure*}

One major motivation for using the \textit{STEREO} spacecraft for finding the streamer wave events is to use the coronagraphic data taken from three different viewing angles. Looking at column 4 in Table~\ref{tab:events}, we see that only one event is observed from three different vantage points: streamer wave event number 19 on 2014-04-18. Further on in this paper we will refer to the events by their respective numbers in column 1 of Table~\ref{tab:events}. If we look at the configuration of the spacecraft for event 19 more closely, we can explain why for this particular event there is a higher chance that all three coronagraphs would observe the wave. From Figure~\ref{fig:positions} we can see that \textit{STEREO A} and \textit{B} are close to each other (separation angle of 39\degr{}). \textit{SOHO}, which is positioned at the L1 point between the Earth and the Sun, is at this time opposite to the \textit{STEREO} spacecraft with respect of the Sun, and will thus have the nearly mirrored view of the event in comparison with either of the \textit{STEREO} spacecraft. From the plot in Figure~\ref{fig:positions} one can infer that the streamer slab for this event should be located above the west limb of the Sun as seen from \textit{SOHO}, with a longitudinal extent of around or above 39\degr{}.

From our list of 22 events, 11 were observed from 2 different viewpoints. The separation angle between the two observing spacecraft ranges from 118\degr{} to 175\degr{} for these events. The average of the separation angles is 144.5\degr{}. These values indicate that each time an event is observed from two different viewpoints, they have a quite high separation angle. This often means that the two observations are close to each other's mirror image. The downside from this is that the nearly mirror image does not give much extra information that would allow to derive the three-dimensional (3D) configuration of the streamer (except perhaps the longitudinal extent). All 10 other streamer wave events are only observed by one of the three coronagraphs. We do have to note here that for the last two events \textit{STEREO B} data was not available, so we can not exclude that this event was also visible from this angle.

From Figure~\ref{fig:positions} one can get a good idea of how the streamer slab should be positioned for each event and the slab longitudinal extent, especially if the streamer is observed by more than one coronagraph. For 4 events, the 3D picture is somewhat less clear. \edit1{We also expected to observe} events 4, 5, and 7\edit1{\deleted{would be expected to also be observed}} from \textit{STEREO B}/COR2 as the separation of the \textit{STEREO} spacecraft was close to 180\degr{} (see Figure~\ref{fig:positions})\edit1{\added{, but this was not the case}}. For event 7, the signal-to-noise ratio of \textit{STEREO B}/COR2 images during the event is noticeably low. The streamer and the oscillation seem to be present, but the data quality does not allow us to make reliable measurements. Events 4 and 5 are more difficult to explain, but we noticed that the streamer is a significantly weaker structure in \textit{STEREO B}/COR2 than in \textit{STEREO A}/COR2, while one would expect a similar intensity. Together with the low amplitude of the waves for these two events, they could not be detected due to a too low signal. \edit1{\added{In the running difference images of \textit{STEREO B}/COR2, the signal-to-noise ratio of the wave feature was too weak to mark it as a visual detection in accordance to the procedure outlined in the beginning of this Section.}} The fourth unclear event is event 10. From the plot in Figure~\ref{fig:positions} one can hardly imagine the position of one streamer slab such that it is visible from both \textit{STEREO B}/COR2 and \textit{SOHO}/LASCO C2+C3, but not from \textit{STEREO A}/COR2. A possible explanation could be that the two coronagraphs observed two different streamers each confined to a narrow longitude range near the plane of the sky of the respective coronagraph. Event 10 is the event with the lowest separation angle in our data set (118\degr{}). If there indeed are two different streamers, then the lowest separation angle between two spacecraft that simultaneously observed an event becomes 127\degr{}. 

Another peculiar event that we would like to note here, is event 12. The position angle (PA) of the streamer at 5 $R_{\sun}$ (measured counterclockwise from the solar north) for this event is 0\degr{}. From the configuration of the spacecraft near the quadrature at that time (see Figure~\ref{fig:positions}) one can derive that the streamer slab should be situated approximately along the Sun-\textit{STEREO B} line above the north pole, so \textit{STEREO A}/COR2 should have a face-on view of it. Unfortunately, from the \textit{STEREO A} perspective the CME passes right in front of the slab during the event, which obstructs the view of the streamer slab during the event.

The observations by different coronagraphs in our data set indicate that one needs a very specific angle with respect to the streamer to be able to see the streamer wave event. \edit1{\added{A possible explanation is the alignment of the line of sight of the coronagraph with the streamer slab. When those are well-aligned, the oscillation is situated in (or close to) the plane of the sky of the coronagraph. The observed integrated brightness is then not changed much (by the misalignment between the slab and the coronagraph plane of the sky) and the whole slab can be seen oscillating as a single propagating wave going through a narrow streamer.}} One could investigate this relation more closely through reconstructing the 3D structure of the corona for each event separately \citep[see e.g. ][]{decraemer_2019} and analyzing why one can only observe the streamer wave events from specific viewing angles. The events found during the periods around the quadratures between \textit{SOHO} and \textit{STEREO A/B} in 2011 and between \textit{STEREO A} and \textit{B} in 2013 could be especially interesting for this. 

\begin{rotatetable*}
\begin{deluxetable*}{cccccccccccc}

\centerwidetable

\tablecaption{Summary of the physical parameters measured for the 22 streamer wave events identified with \textit{STEREO}/COR2.\label{tab:events}}

\tablehead{\multicolumn{2}{c}{} & \multicolumn{6}{c}{Wave} & \multicolumn{4}{c}{CME} \\ \cmidrule(l{2pt}r{2pt}){3-8} \cmidrule(l{2pt}r{2pt}){9-12} \colhead{Number} & \colhead{Date} & \colhead{Time} & \colhead{Observed by} & \colhead{PA} & \colhead{Period} & \colhead{Wavelength} & \colhead{Phase speed} & \colhead{Time} & \colhead{CPA} & \colhead{Width} & \colhead{Speed} \\ 
\colhead{} & \colhead{} & \colhead{(UT)} & \colhead{} & \colhead{(\degr{})} & \colhead{(min)} & \colhead{($\mathrm{R}_{\odot}$)} & \colhead{($\mathrm{km \ s}^{-1}$)} & \colhead{(UT)} & \colhead{(\degr{})} & \colhead{(\degr{})} & \colhead{($\mathrm{km \ s}^{-1}$)} }

\startdata
1 & 2008-04-05 & 16:22:00 & A & 236 & 120 & 6.00 & 670 & 16:07:30 & 277 & 45 & 1140 \\
2 & 2008-05-17 & 10:37:30 & A & 121 & 150 & 4.83 & 680 & 10:37:30 & 96 & 30 & 930 \\
3 & 2011-04-07 & 12:08:00 & A (175\degr{}) B & 277 & 360 & 6.25 & 560 & 12:08:00 & 221 & 66 & 860 \\
4 & 2011-04-27 & 03:08:15 & A & 132 & 240 & 5.80 & 520 & 02:54:00 & 52 & 47 & 650 \\
5 & 2011-06-04 & 22:54:00 & A & 137 & 165 & 5.13 & 480 & 22:08:00 & 292 & 360 & 2030 \\
6 & 2011-06-20 & 18:24:00 & A (172\degr{}) B & 250 & 300 & 5.58 & 420 & 18:08:15 & 284 & 42 & 520 \\
7 & 2011-08-04 & 04:54:00 & A & 325 & 480 & 7.16 & 740 & 04:08:15 & 58 & 100 & 1700 \\
8 & 2012-03-15 & 08:39:00 & A & 140 & 180 & 4.24 & 640 & 08:39:00 & 81 & 55 & 210 \\
9 & 2012-05-12 & 00:24:00 & A (128\degr{}) B & 143 & 210 & 4.23 & 490 & 23:54:00 & 82 & 72 & 620 \\
10 & 2012-05-17 & 02:39:21 & B (118\degr{}) C2+C3 & 219 & 180 & 6.07 & 730 & 02:24:00 & 262 & 199 & 670 \\
11 & 2012-06-14 & 14:54:00 & A (127\degr{}) B & 17 & 330 & 7.09 & 610 & 13:54:00 & 116 & 113 & 1110 \\
12 & 2013-01-15 & 20:39:58 & B & 0 & 270 & 6.86 & 740 & 20:09:13 & 42 & 39 & 970 \\
13 & 2013-02-06 & 00:39:00 & A & 90 & 150 & 4.12 & 570 & 00:39:00 & 37 & 50 & 940 \\
14 & 2013-03-13 & 00:39:00 & A & 210 & 255 & 2.86 & 470 & 00:39:00 & 255 & 45 & 630 \\
15 & 2013-05-01 & 03:08:15 & A (135\degr{}) C2+C3 & 233 & 300 & 6.93 & 730 & 02:54:00 & 283 & 68 & 930 \\
16 & 2013-05-01 & 03:39:00 & A (135\degr{}) C2+C3 & 11 & 270 & 5.35 & 590 & 02:54:00 & 283 & 68 & 930 \\
17 & 2013-05-14 & 23:24:00 & A (136\degr{}) C2+C3 & 11 & 225 & 6.61 & 670 & 23:08:15 & 330 & 31 & 900 \\
18 & 2013-05-27 & 20:08:15 & A & 175 & 195 & 5.92 & 490 & 19:39:00 & 220 & 47 & 510 \\
19 & 2014-04-18 & 14:08:15 & A (39\degr{}) B (165\degr{}) C2+C3 & 99 & 150 & 2.99 & 360 & 13:08:15 & 117 & 122 & 960 \\
20 & 2014-06-05 & 16:08:36 & B (164\degr{}) C2 & 40 & 300 & 5.10 & 430 & 15:24:21 & 101 & 71 & 790 \\
21 & 2015-12-11 & 05:24:00 & A (167\degr{}) C2+C3 & 249 & 135 & 4.38 & 640 & 05:08:15 & 268 & 21 & 680 \\
22 & 2017-07-14 & 02:24:00 & A (133\degr{}) C2+C3 & 257 & 300 & 8.55 & 720 & 01:54:00 & 172 & 246 & 1000 \\
\enddata

\tablecomments{The first column shows the number we assigned to each event. The second to eighth columns give the date, the starting time of the observation, the coronagraphs that observed the wave (the separation angle between the spacecraft is given in brackets), the position angle (PA) of the streamer axis at 5 $\mathrm{R}_{\odot}$, the period, the wavelength, and the phase speed of the wave. The first coronagraph mentioned in the fourth column is considered to be the main coronagraph for each event. The last four columns present the starting time of the CME, the CME's central position angle (CPA), angular width, and plane-of-the-sky radial speed. The parameters of the CME are measured in the field of view (FOV) of the main coronagraph.
(This table will be available in a machine-readable version.)}

\end{deluxetable*}
\end{rotatetable*} 

Visibility only from a specific angle could also help to explain why we observe these events so rarely. We found only 22 events in a span of almost 9 years, while there are much more interactions reported between CMEs and streamers. Sometimes the interaction is present, but does not result in a streamer wave event according to our criteria. For example, the streamer is displaced but does not move back and thus shows only 1 black/white patch in the running difference images. On the other hand, only 8 streamer wave events were observed by \citet{feng_streamer_2011} during the solar cycle 23. One can also notice that significantly more events were observed with \textit{STEREO A}/COR2 than with \textit{STEREO B}/COR2 (respectively 19 for \textit{A} versus 8 for \textit{B}). This is comparable to the 8 events reported by \citet{feng_streamer_2011}, since they were found in an about equally long period during solar cycle 23. Then, it seems that \textit{SOHO}/LASCO \edit1{\added{C2+C3}} and \textit{STEREO B}/COR2 are about equally good at detecting streamer wave events. \textit{STEREO A}/COR2 however found around a double number of events. This could be due to higher resolution of COR2 compared to LASCO C3. The lower signal-to-noise ratio in \textit{STEREO B}/COR2 could then be the reason why it only found a number of events comparable to that of LASCO C3 and less than that of \textit{STEREO A}. The higher number of events could also be due to our survey of the whole \textit{STEREO} data set, instead of restricting ourselves to time periods around CMEs reported in a given catalog, as was done by \citet{feng_streamer_2011}.

\begin{figure*}
\centering
\figurenum{\ref{fig:positions} (continued)}
\includegraphics[width=0.31\textwidth]{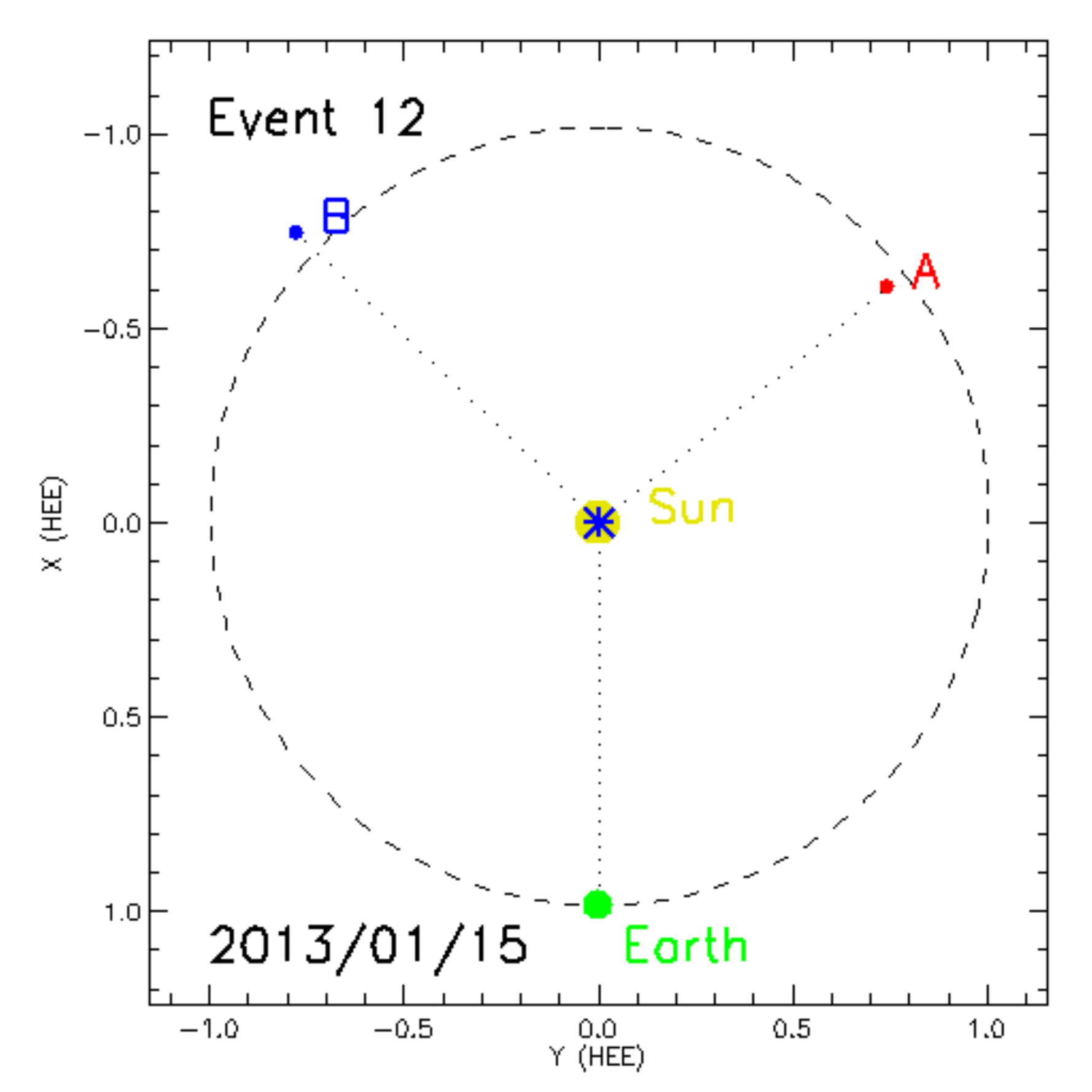}~
\includegraphics[width=0.31\textwidth]{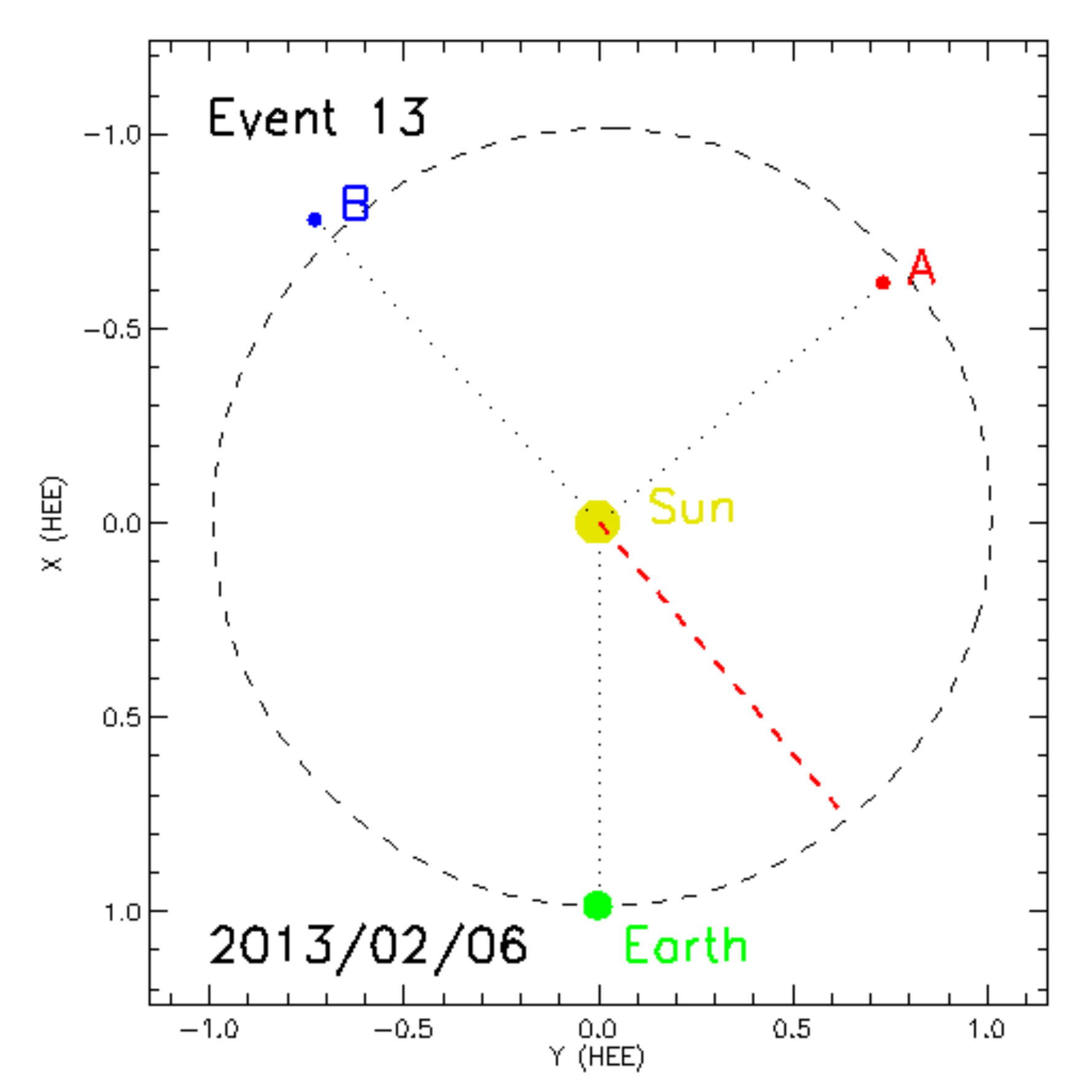}~
\includegraphics[width=0.31\textwidth]{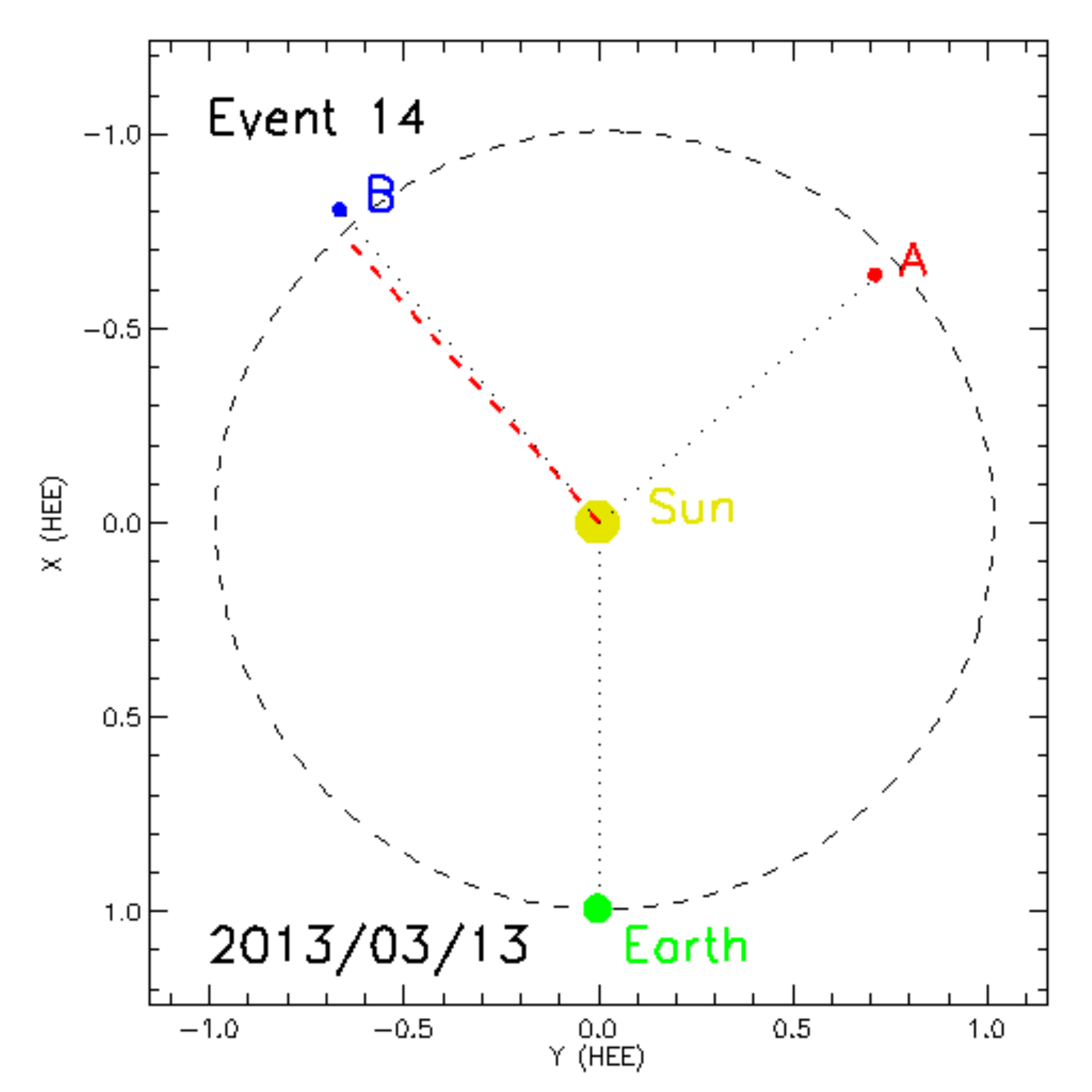}\\
\includegraphics[width=0.31\textwidth]{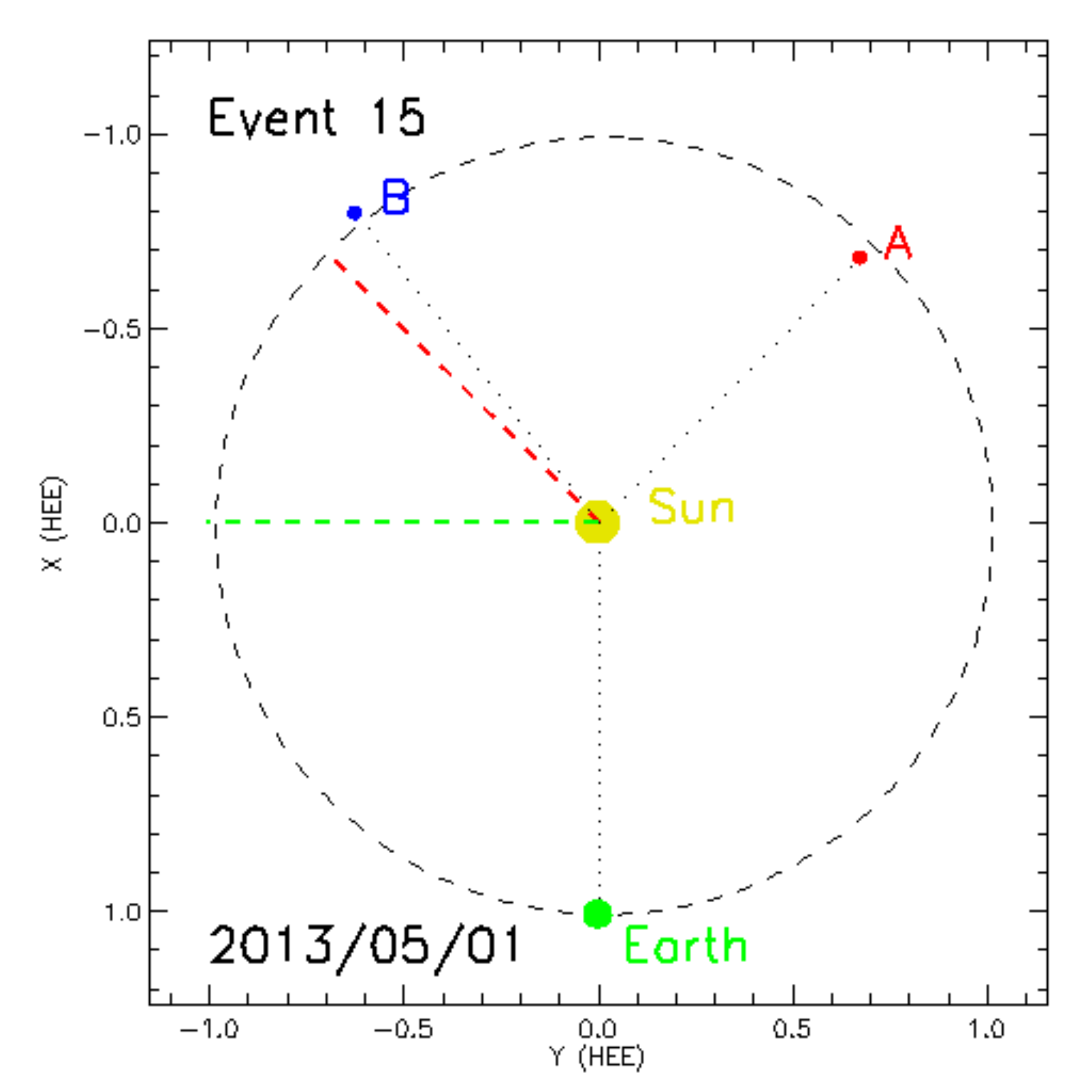}~
\includegraphics[width=0.31\textwidth]{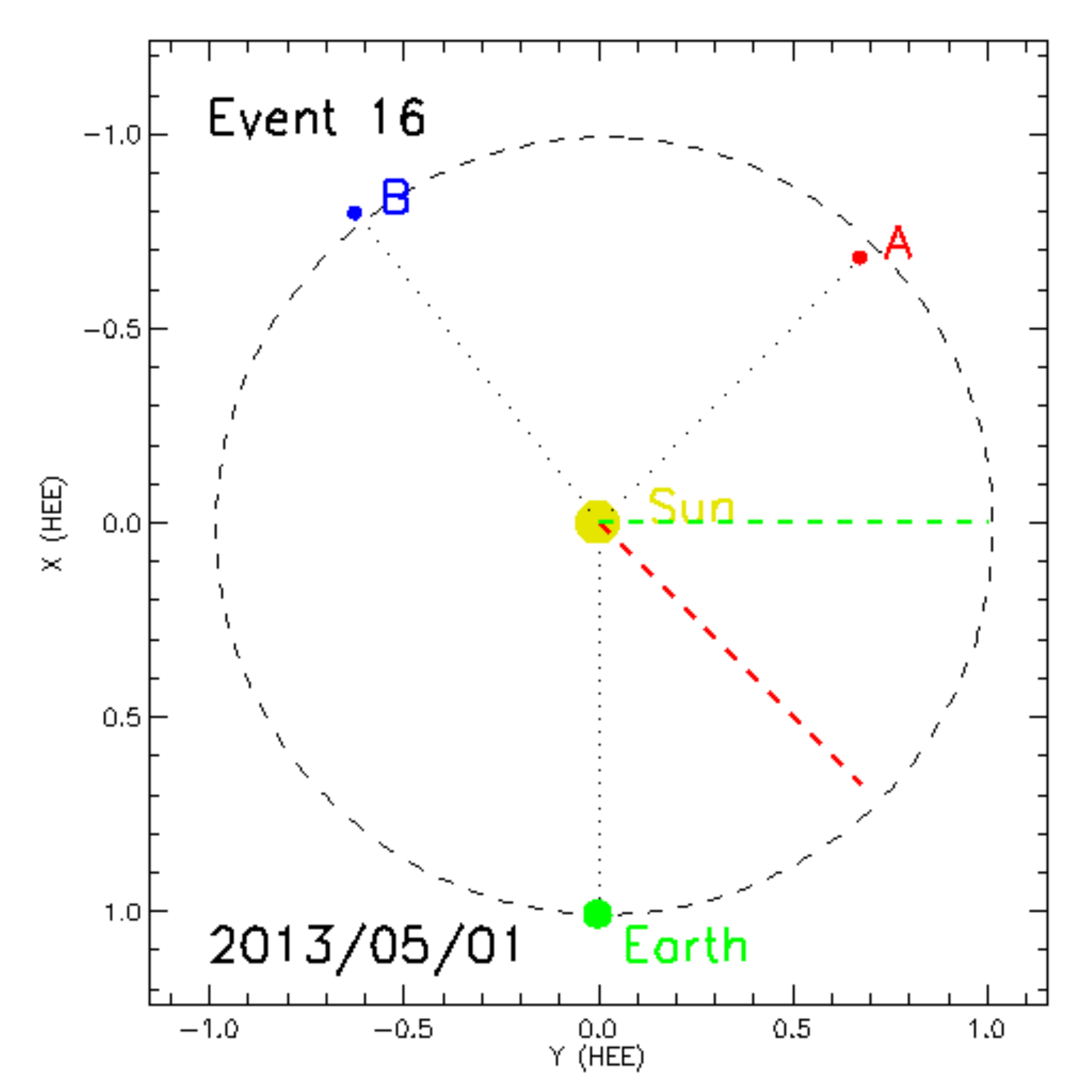}~
\includegraphics[width=0.31\textwidth]{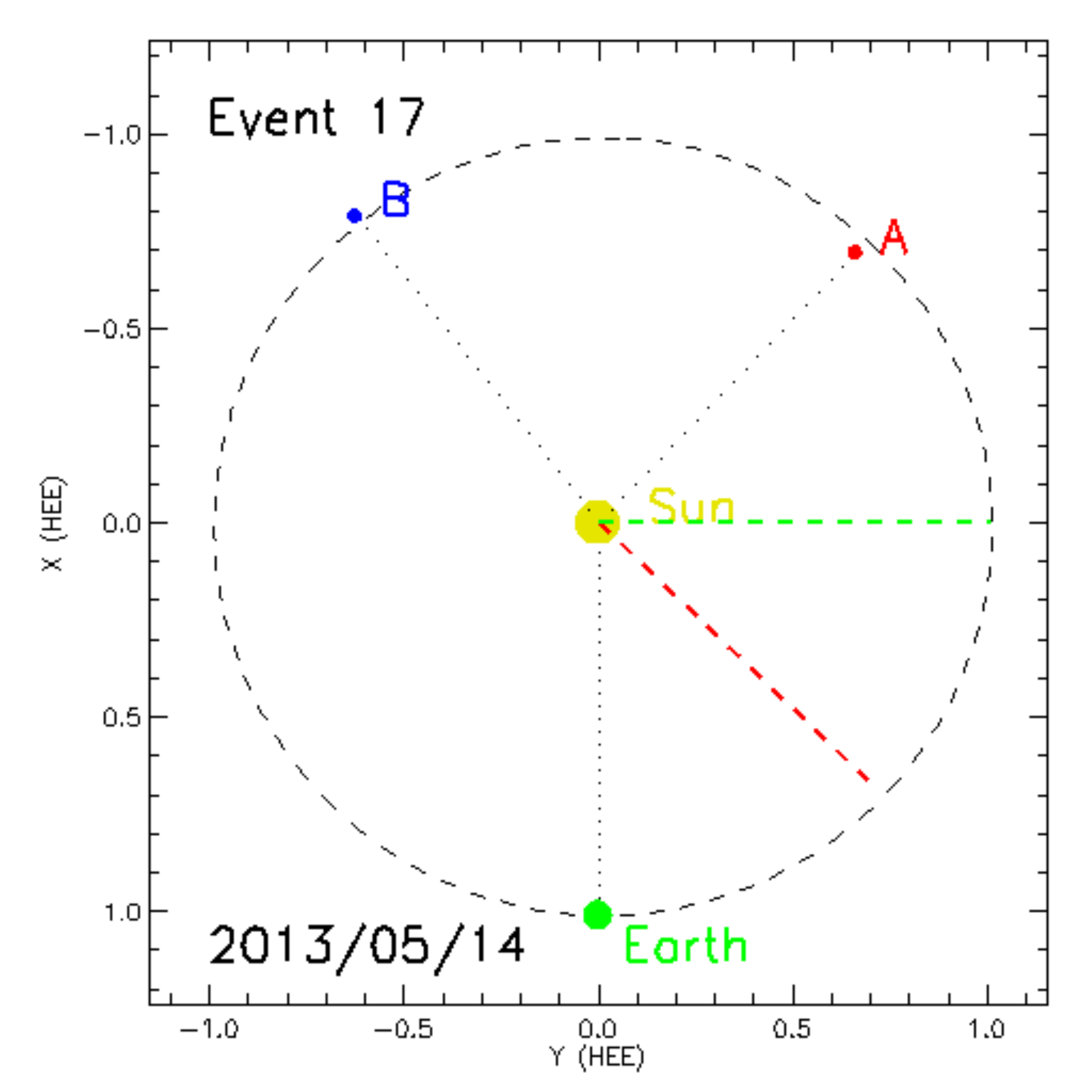}\\
\includegraphics[width=0.31\textwidth]{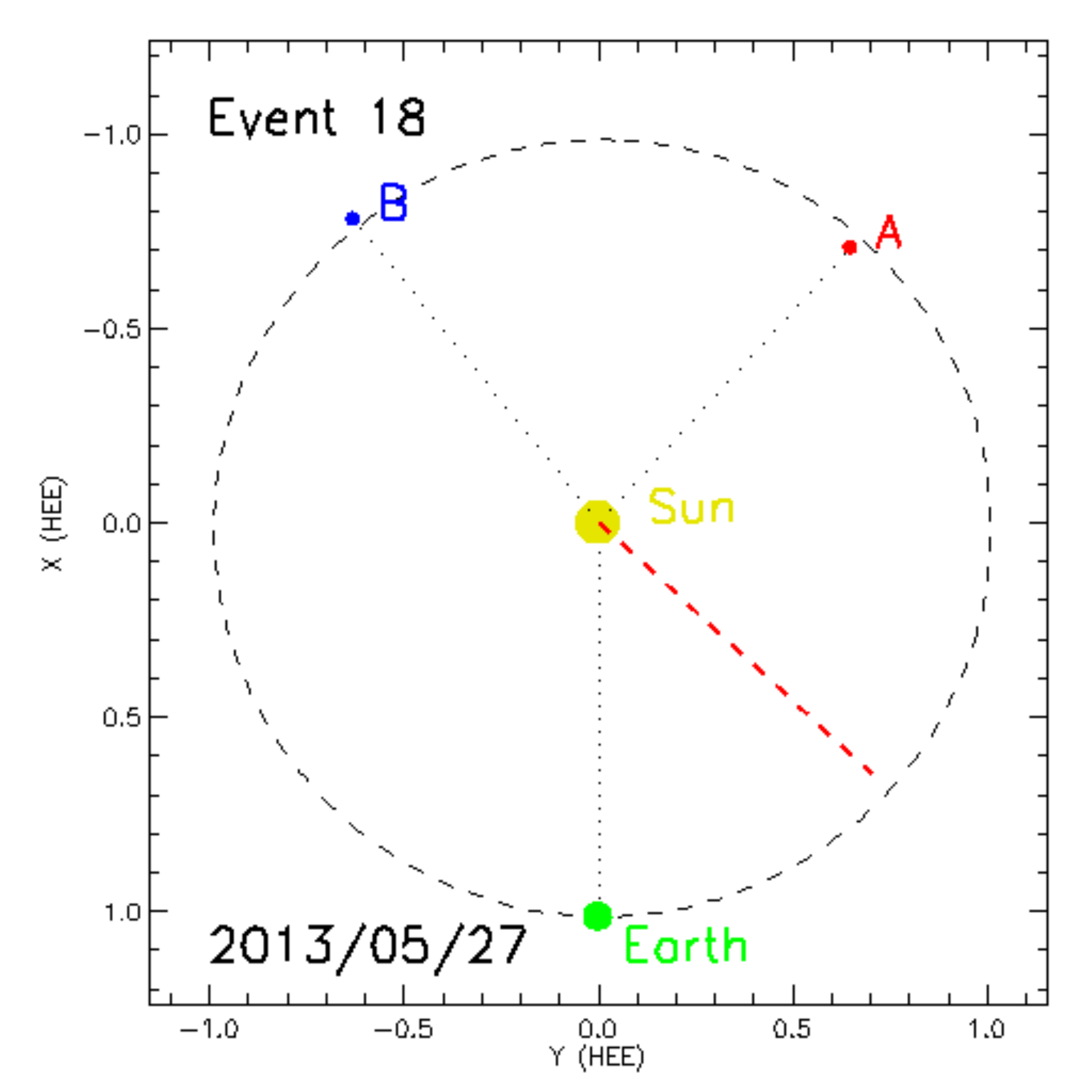}~
\includegraphics[width=0.31\textwidth]{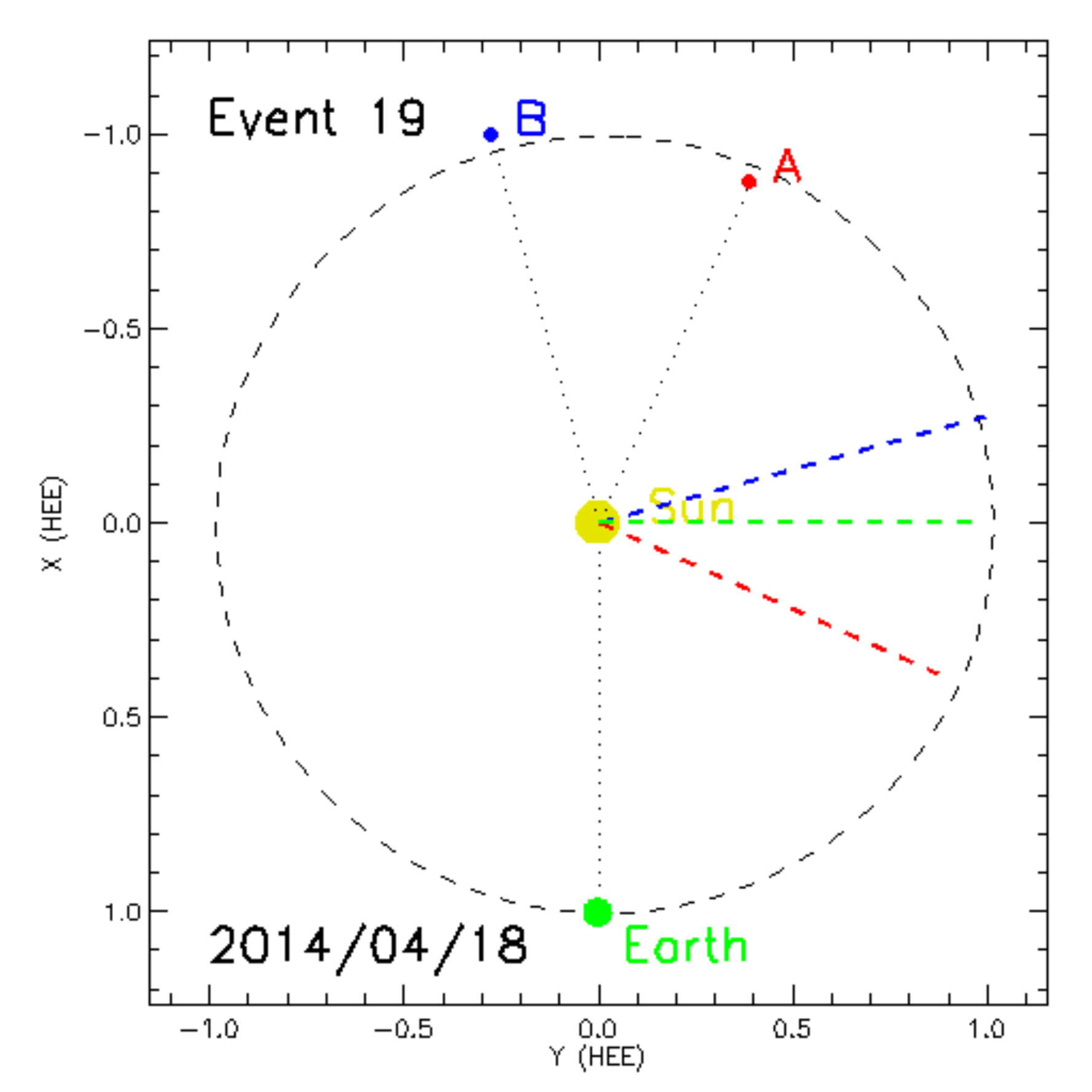}~
\includegraphics[width=0.31\textwidth]{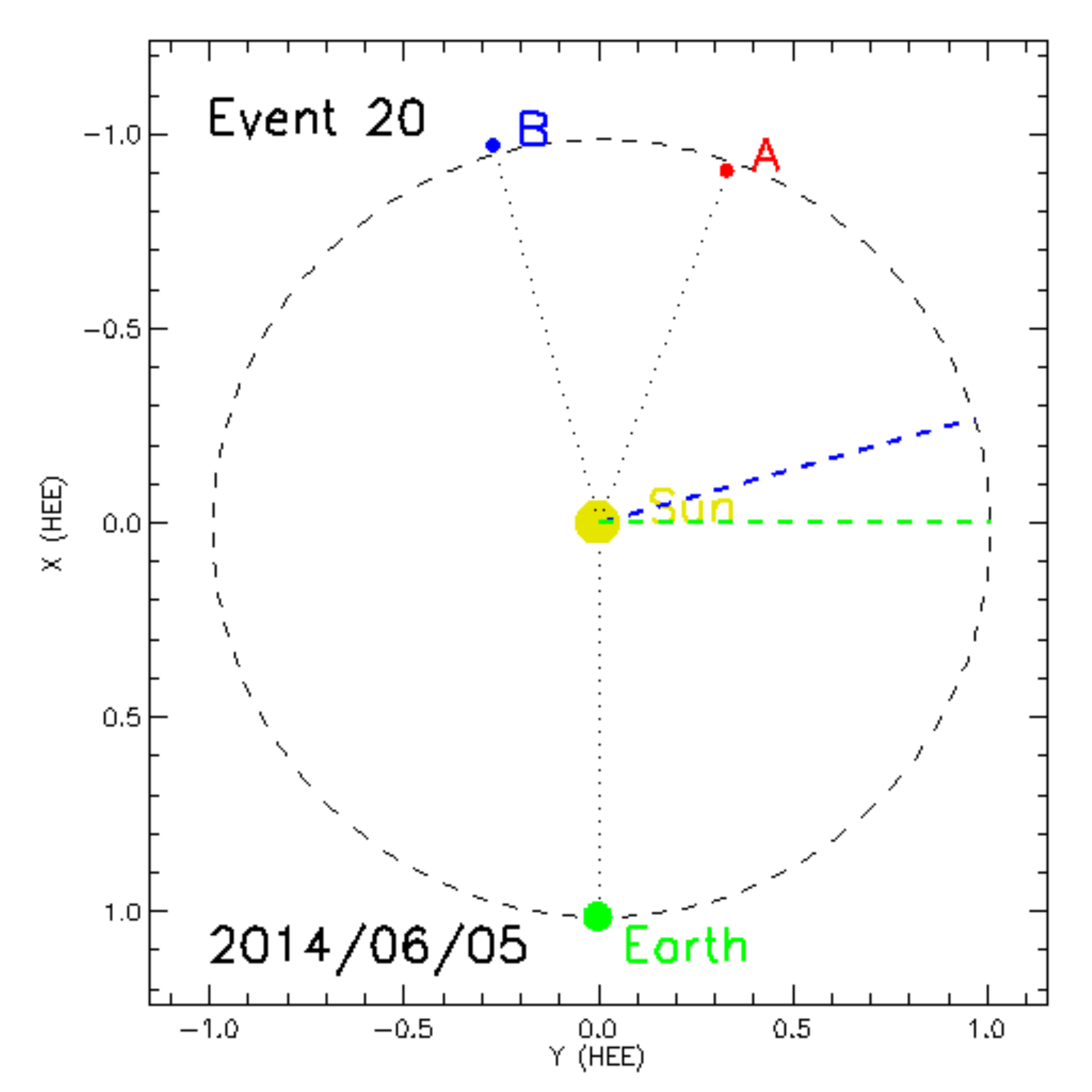}\\
\includegraphics[width=0.31\textwidth]{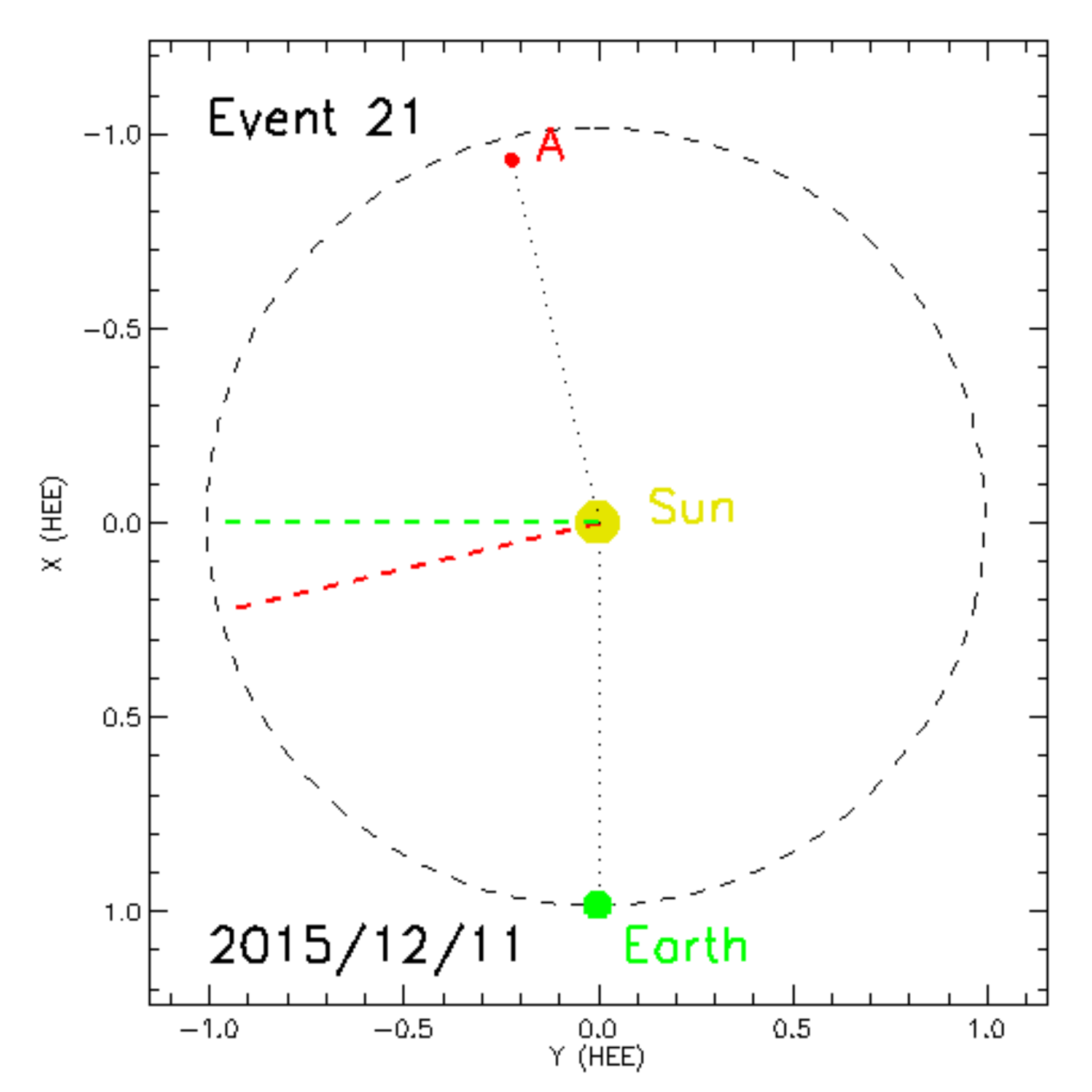}~
\includegraphics[width=0.31\textwidth]{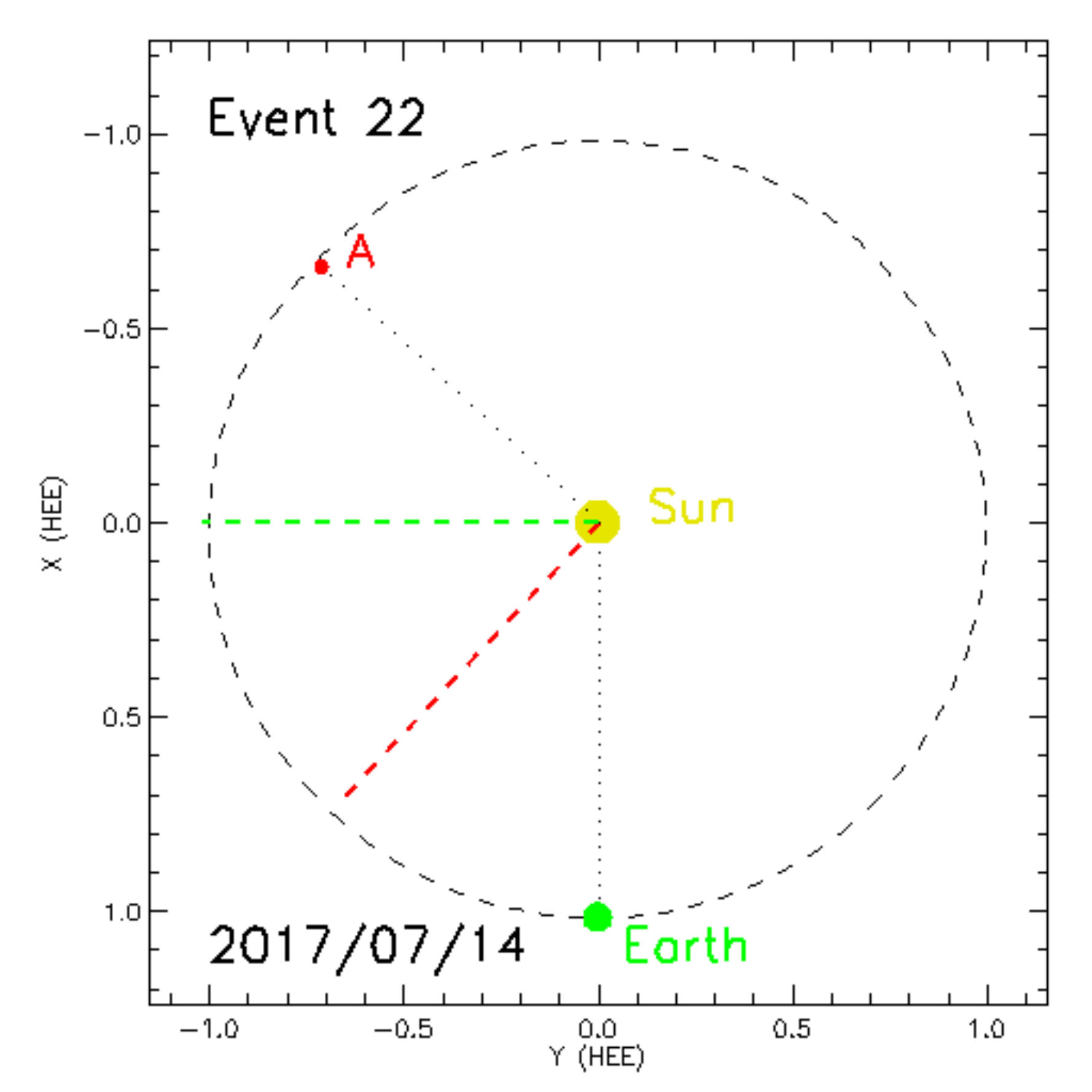}
\caption{}
\end{figure*}

\begin{figure}
\centering
\includegraphics[width=0.45\textwidth]{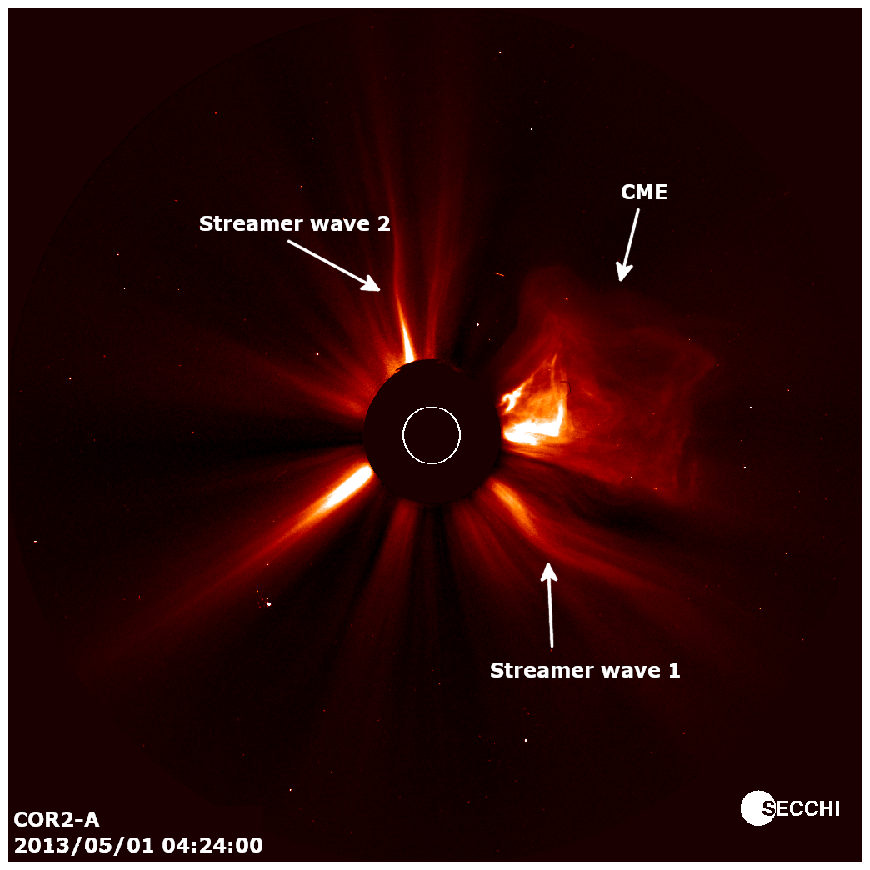}
\caption{\textit{STEREO A}/COR2 white-light image of the two streamer wave events 15 and 16 on 2013-05-01. The CME perturbs two different streamers and excites two different streamer wave events.}
\label{fig:2streamer1cme}
\end{figure}

In this section we also briefly comment on five events. \edit1{\added{These five events behave somewhat differently than the other 17 ``typical'' events. With a typical event, we mean that there is one clear CME initiating an oscillation in one streamer in the FOV of the main coronagraph (see Section~\ref{s:measure}).}} The first two \edit1{\added{``atypical'' events}} are actually a pair of streamer wave events (events 15 and 16) on 2013-05-01 that we highlight since they have the same initiating CME. This is the only such occasion in our data set. The events are shown in Figure~\ref{fig:2streamer1cme}. The CME is first visible in the COR2 A field of view (FOV) at 02:54:00 UT and originates just above the solar equator on the west limb. It then disturbs two different streamers and initiates a streamer wave event in both of them. The wave in the southern streamer at a PA of 233\degr{} starts a little earlier than the wave propagating in the northern streamer located at a PA of 11\degr{}. This can be explained by the fact that the CME originates from a location that appears to be closer to the southern streamer. 

\begin{figure*}
\centering
\includegraphics[width=\textwidth]{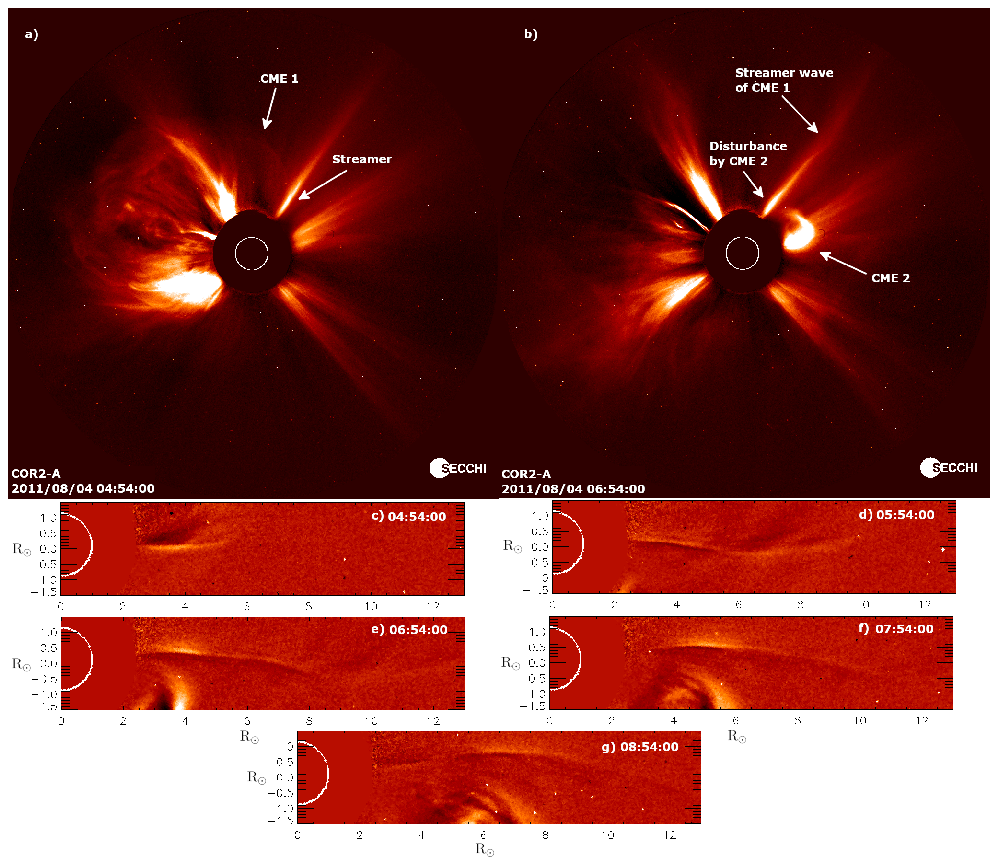}
\caption{\textit{STEREO A}/COR2 white-light images of streamer wave event 7 on 2011-08-04 (panels (a) and (b)). The streamer is perturbed for the second time by CME 2 after the excitation of the streamer wave event by CME 1. Panels (c) to (g) show running difference images of the streamer rotated clockwise by 55\degr{} at 5 times during the event.}
\label{fig:1streamer2cme}
\end{figure*}

The two events 7 and 17 have a special feature that is shown for event 7 in Figure~\ref{fig:1streamer2cme}. After the CME perturbing the streamer and the start of \edit1{\added{the}} wave, a second CME is launched\edit1{\added{,}} which also perturbs the streamer and interacts with the ongoing oscillation. This makes these two events much more difficult to analyze since it is very hard to disentangle which effects are caused by each of the two incident CMEs. It could be that event 13 also fits in this category\edit1{\added{, but we will elaborate on this in the next Section}}.

Event 7 is also interesting for another reason. \citet{kwon_2013} reported an observation of an oscillation of a coronal streamer\edit1{\added{, located at a PA of about 30\degr{},}} with \textit{STEREO} COR1\edit1{\added{ at approximately the same time as event 7 occurred. Since event 7 is located at a PA of 325\degr{} in the COR2 FOV, it can however not be the same streamer}}. The streamer in which \edit1{\replaced{they}{\citet{kwon_2013}}} see the oscillation, gets completely disrupted by the CME \edit1{in the COR2 FOV} and \edit1{\added{causes the streamer to}} disappear \edit1{\added{after the disruption}}. \edit1{\replaced{They}{\citet{kwon_2013}}} did identify the streamer \edit1{of event 7} \edit1{\added{located at a PA of around 330\degr{} in the COR1 FOV}} (S2 in \edit1{\added{their}} Figure 1)\edit1{\replaced{and}{, but only}} mention\edit1{\added{ed}} that it gets deflected. \edit1{\added{They could track an almost coinciding coronal disturbance in COR1 and EUVI images originating from the CME source region. The coronal disturbance that they have identified to be a fast magnetosonic wave, passes through the streamers, which results in the deflection and oscillation of the streamers. \citet{kwon_2013} did not discuss in detail the interaction between the disturbance and the streamers.}} It would be interesting to connect these two observations to get a more complete picture of this streamer wave event. 

The final event that we want to comment on is event 14. For this event there are actually two CMEs which come close to the streamer at the initiating time of the streamer wave event. From the coronagraph images it is almost impossible to clearly see which of the two is the CME that actually perturbs the streamer and initiates the streamer wave event. In our study here we assume the CME that originated closest to the streamer location in PA to be the initiating CME.

\section{Measured properties of streamer wave events}\label{s:measure}
For each individual event, we measured some main properties. We illustrate our measurements using one of our typical events, event 13, as an example. First, the observations by all available coronagraphs were examined to see if the wave was visible from multiple viewpoints. We then selected the coronagraph in which the streamer wave was most clearly visible as the main coronagraph. If the wave was clearly visible in more than one coronagraph, we selected the coronagraph with the highest resolution and/or highest signal-to-noise ratio as the main coronagraph (thus favoring COR2 over LASCO C3, and COR2 A over COR2 B). We did not use LASCO C2 as a main coronagraph, even though it has the highest resolution, since the wavelength of several streamer wave events exceeds the C2 FOV between 2.2 and 6 $\mathrm{R}_{\odot}$. \edit1{\added{Essentially this means that all our measurements were taken in one of the two COR2 coronagraphs.}} In event 20, the streamer wave event was only observed by LASCO C2 and not by LASCO C3 because the pylon of the occulting disk obscured the location of the streamer in the LASCO C3 FOV. For each event the observing coronagraphs can be found in the fourth column of Table \ref{tab:events}, with the main coronagraph listed first.

Next, we determine the position angle (PA) of the streamer axis in the FOV of the main coronagraph, measured counterclockwise from the solar north. We extract a circular brightness profile at 5 $\mathrm{R}_{\odot}$ and we locate the peak of brightness that corresponds to the streamer. For event 13, this gives a position angle of 90\degr{}\edit1{\added{, as can be seen in Figure~\ref{fig:location}}}. This procedure is done for the image right before the streamer event takes place. The PA of the streamer axis for each event can be found in column 5 of Table~\ref{tab:events}.

\edit1{
\begin{figure*}
\centering
\includegraphics[width=\textwidth]{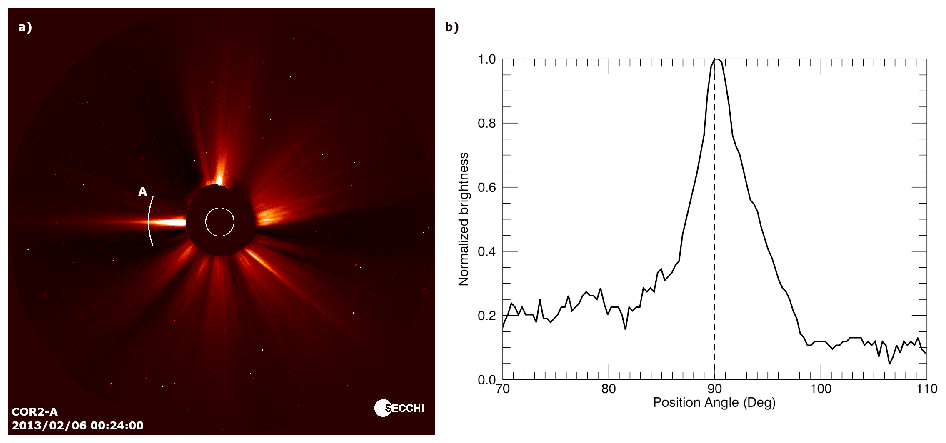}
\caption{Panel (a) shows the white-light \textit{STEREO} A/COR2 image taken on 2013-02-06 at 00:24:00, with arc A overplotted at 5 $\mathrm{R}_{\odot}$. Panel (b) shows the normalized profile of brightness at 5 $\mathrm{R}_{\odot}$ (along the arc shown in panel (a)). The dashed line in panel (b) indicates the identified PA of the streamer axis at 90\degr{}.}
\label{fig:location}
\end{figure*}
}

\begin{figure*}
\centering
\includegraphics[width=\textwidth]{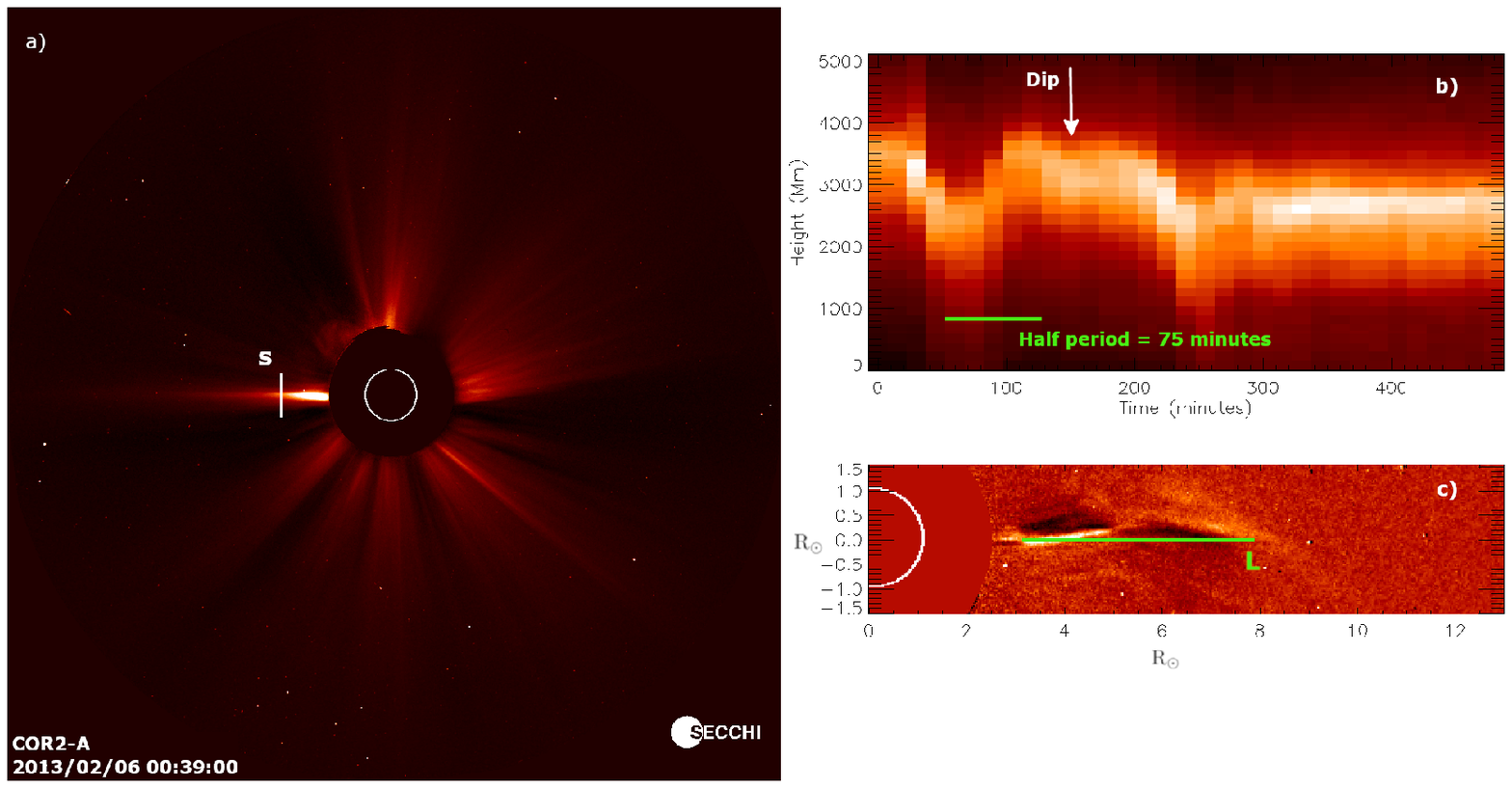}
\caption{\textit{STEREO} A/COR2 white-light image of the streamer of event 13 on 2013-02-06 (panel (a)) and in panel (b) the time-distance map along the slit S shown in panel (a). The green line in panel (b) indicates how we performed the period measurement of 150 minutes. Panel (c) shows the running difference image of the streamer wave event on 2013-02-06 at 01:54:00 UT, rotated clockwise by 180\degr{}. The green line L indicates the wavelength as can be measured in this frame (4.12 $\mathrm{R}_{\odot}$).}
\label{fig:timedistance}
\end{figure*}

To measure the period of the streamer wave events, we first created a time-distance map along a slit situated across the streamer stalk, as shown in panel (b) of Figure~\ref{fig:timedistance}. The slit intersects the streamer stalk at 5 $\mathrm{R}_{\odot}$ and has a width of 1 pixel, as can be seen in panel (a) of Figure~\ref{fig:timedistance}. Due to the relatively low cadence (15 minutes) of the COR2 instruments, the time-distance maps have a relatively low resolution in the time axis. The period is determined as the time between two wave crests or troughs on the time-distance map. For some events, there are disturbances in the time-distance map which do not allow the measurement of the complete period. This is the case for event 13, where one can notice a dip (indicated with an arrow in Figure~\ref{fig:timedistance}(b)): it is uncertain if it belongs to the streamer wave event or is an additional disturbance. We do not clearly see a second CME being launched, but the dip in the time-distance map could be due to a second disturbance in the solar corona. Looking at the data from COR2 B and LASCO C2 and C3 from the same period, we notice that two CMEs are clearly distinguishable explaining the disturbance in the time-distance map. Therefore, we believe that event 13 could be similar to events 7 and 17 that were discussed in the previous Section. 

In such cases, we measure the half period between a crest and a trough and calculate the full period from this measurement. For event 13, we can measure a half period of 75 minutes between the first trough and the first crest, which results in a period of 150 minutes. Due to the low resolution in time, we have an error on our period estimations of around 30 minutes (which corresponds at most to 25\%). All period estimates can be found in column 6 of Table~\ref{tab:events}. The time-distance maps also show that the streamer waves decay rapidly. Usually only 1-2 periods are visible on the map. This can be explained by the wave carrying the energy outward in the solar corona\edit1{\replaced{, and}{. The streamer wave propagates through the spherically expanding corona, and thus its initial energy per unit surface decreases due to this geometrical expansion.}} \edit1{\added{The rapid decay}} is also observed in other reports of streamer waves \citep{chen_streamer_2010, feng_streamer_2011, kwon_2013}.

In addition to the period, we measured the wavelength for each streamer wave event. For each event, we chose two frames of running difference images in which a full wavelength is clearly distinguishable. We then manually outlined the length of the wave 2 times in each frame and took the average of these 4 measurements to be the wavelength reported in column 7 of Table~\ref{tab:events}. Panel (c) in Figure~\ref{fig:timedistance} shows the wavelength measurement for the event 13 as an example, which resulted in a wavelength of 4.12 $\mathrm{R}_{\odot}$.

\begin{figure}
\centering
\includegraphics[width=0.45\textwidth]{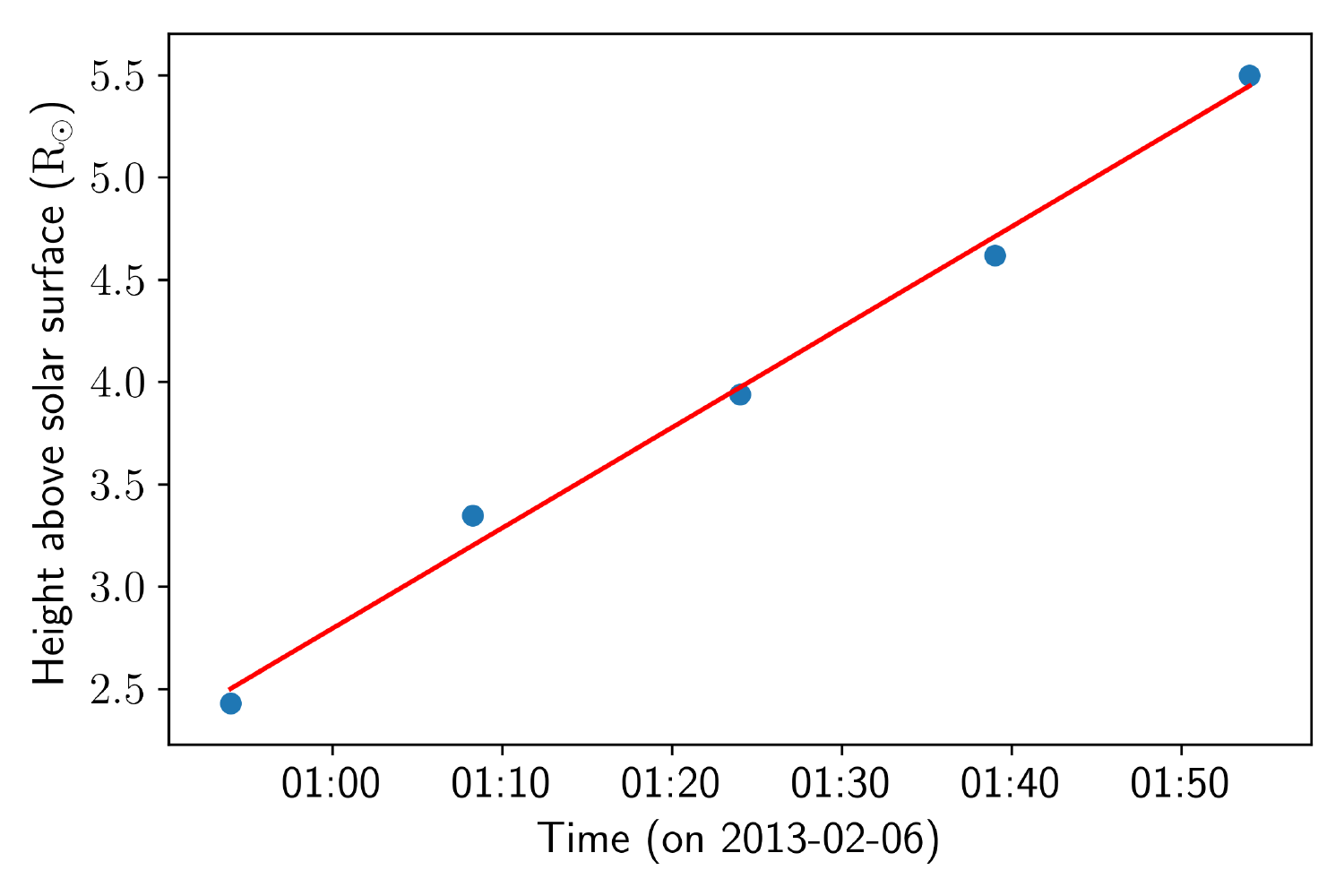}
\caption{Height-time plot of the first crest of the streamer wave event 13 on 2013-02-06. Dots are individual measurements and the red line is the linear fit corresponding to a speed of 570 $\mathrm{km \ s}^{-1}$.}
\label{fig:speed_meas}
\end{figure}

Finally, we also measured the phase speed $v_{ph}$ of the streamer waves. For this, we tracked the height of the first crest or trough of the wave in all \edit1{\replaced{time frames}{images}} that have the wave clearly visible\edit1{\added{, which results in a set of measurement points}}. \edit1{\replaced{Then we calculated}{A linear profile is fitted to these points, from which}} the linear speed in the plane of the sky of the main coronagraph \edit1{\added{is determined, as shown for event 13 in Figure~\ref{fig:speed_meas}}}. \edit1{\added{The measured phase speed of the waves is taken to be equal to this linearly fitted speed.}} For streamer wave event 13, this gives a phase speed of 570 $\mathrm{km \ s}^{-1}$ as can be seen in Figure~\ref{fig:speed_meas}. All measured phase speeds can be found in column 8 of Table~\ref{tab:events}. The error in the speed measurements is around 40 $\mathrm{km \ s}^{-1}$ for the COR2 coronagraphs.
 
Besides measuring the properties of the streamer waves, we also identified different properties of the CMEs that initiated each streamer wave event. The properties of each CME were measured using the main coronagraph identified for its corresponding streamer wave event. The central position angle (CPA), given in column 10 of Table~\ref{tab:events}, of the CME is defined as the midpoint between the PAs of the two flanks of the CME, averaged over all time frames in which the CME is visible. The angular width, given in column 11 of Table~\ref{tab:events}, is measured as the difference between the PAs of the two CME flanks, averaged over all time frames. In column 12 of Table~\ref{tab:events}, the speed of each CME is given. The speeds are measured as the linear plane-of-the-sky speed along the CPA, with a similar height-tracking procedure as for measuring the phase speed of the streamer waves explained above. 

\section{Statistical analysis of the streamer wave events}\label{s:stats}

In this section, we explore the set of streamer events statistically to see if there are any general properties that can be ascribed to this type of events. However, we do recognize that our sample of 22 events is rather small to give any robust statistical evidence. In any case, it is already a large improvement compared to the set of 8 events reported by \citet{feng_streamer_2011}.

\begin{figure*}
\centering
\includegraphics[width=0.45\textwidth]{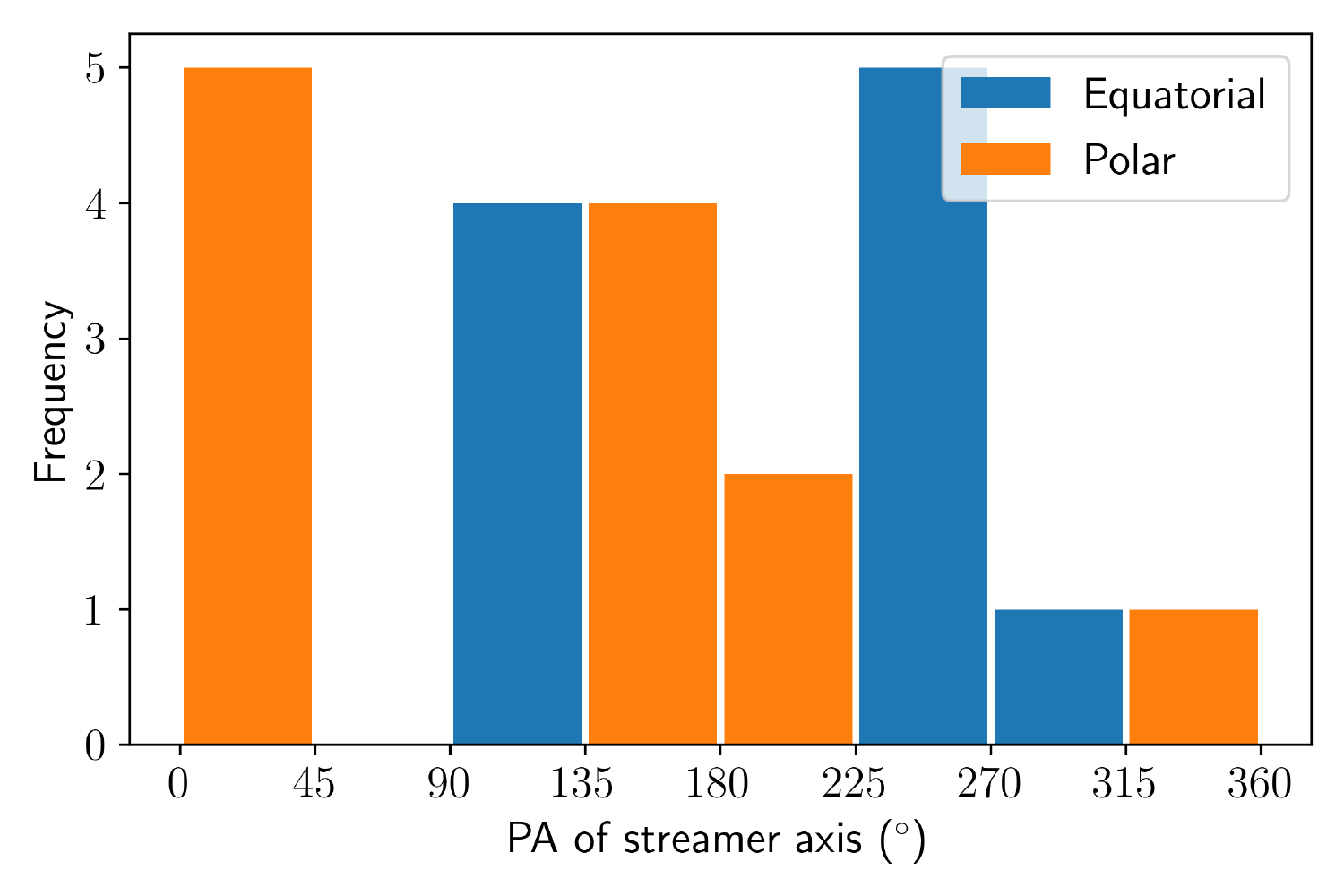}~
\includegraphics[width=0.45\textwidth]{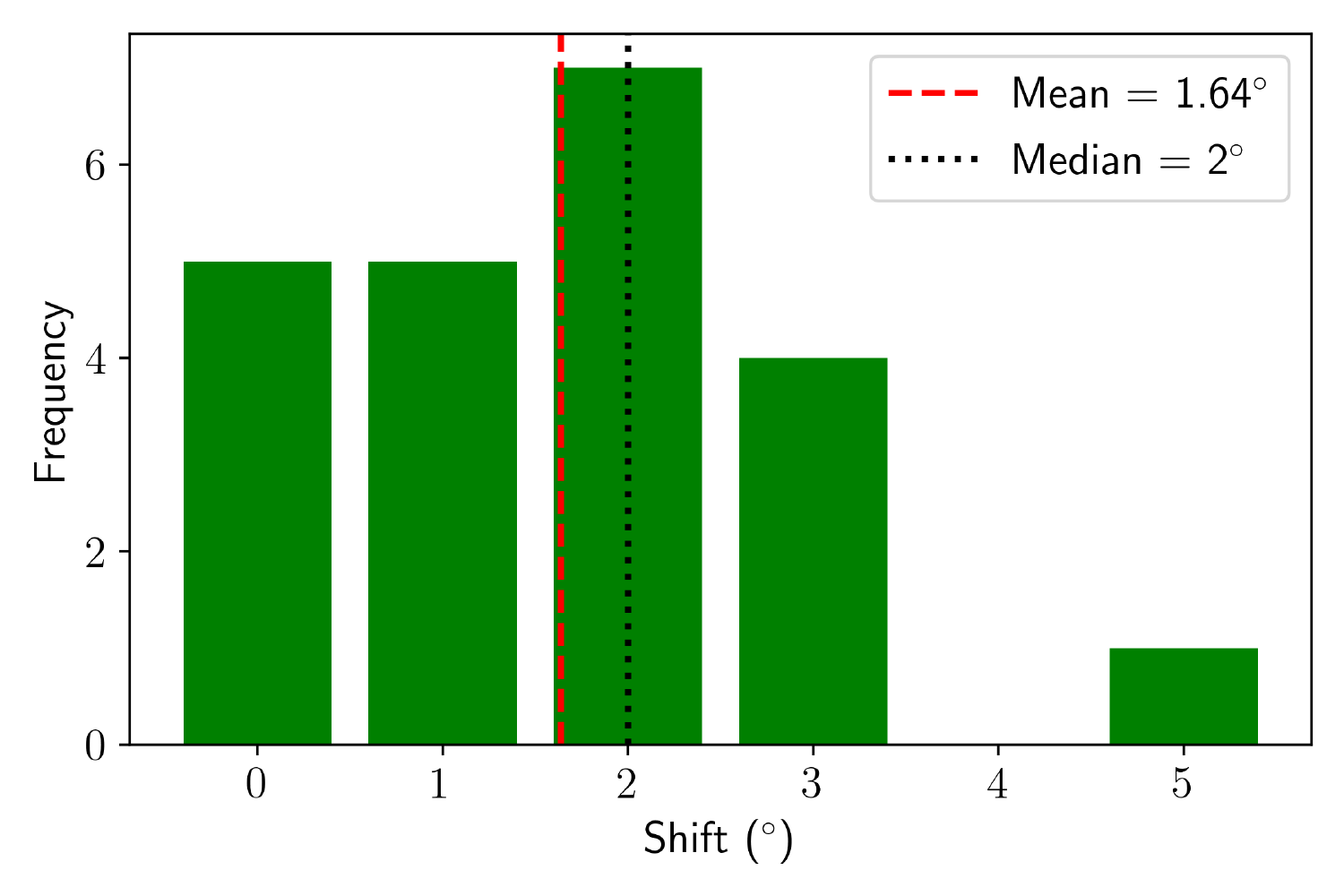}\\
\includegraphics[width=0.45\textwidth]{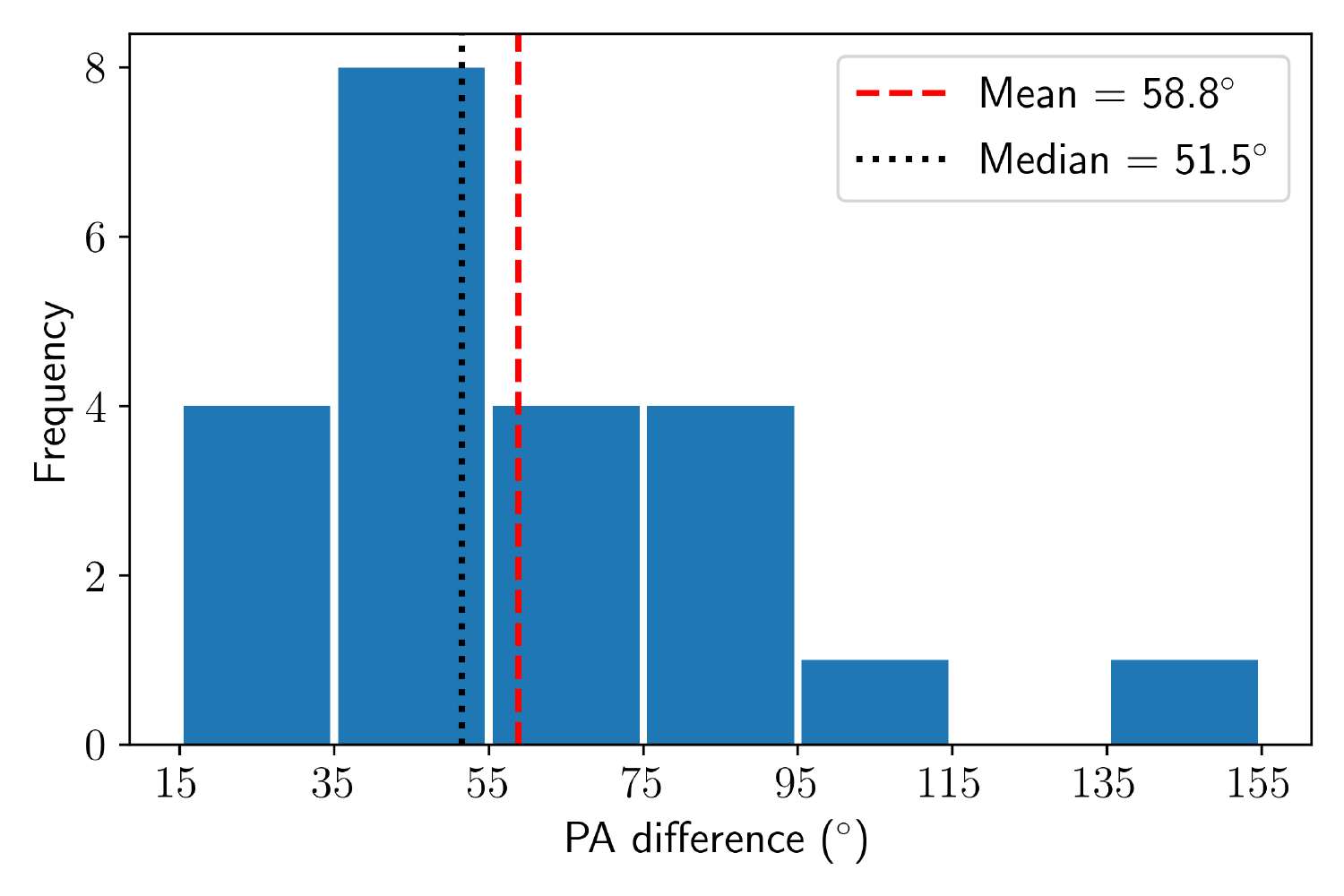}~
\includegraphics[width=0.45\textwidth]{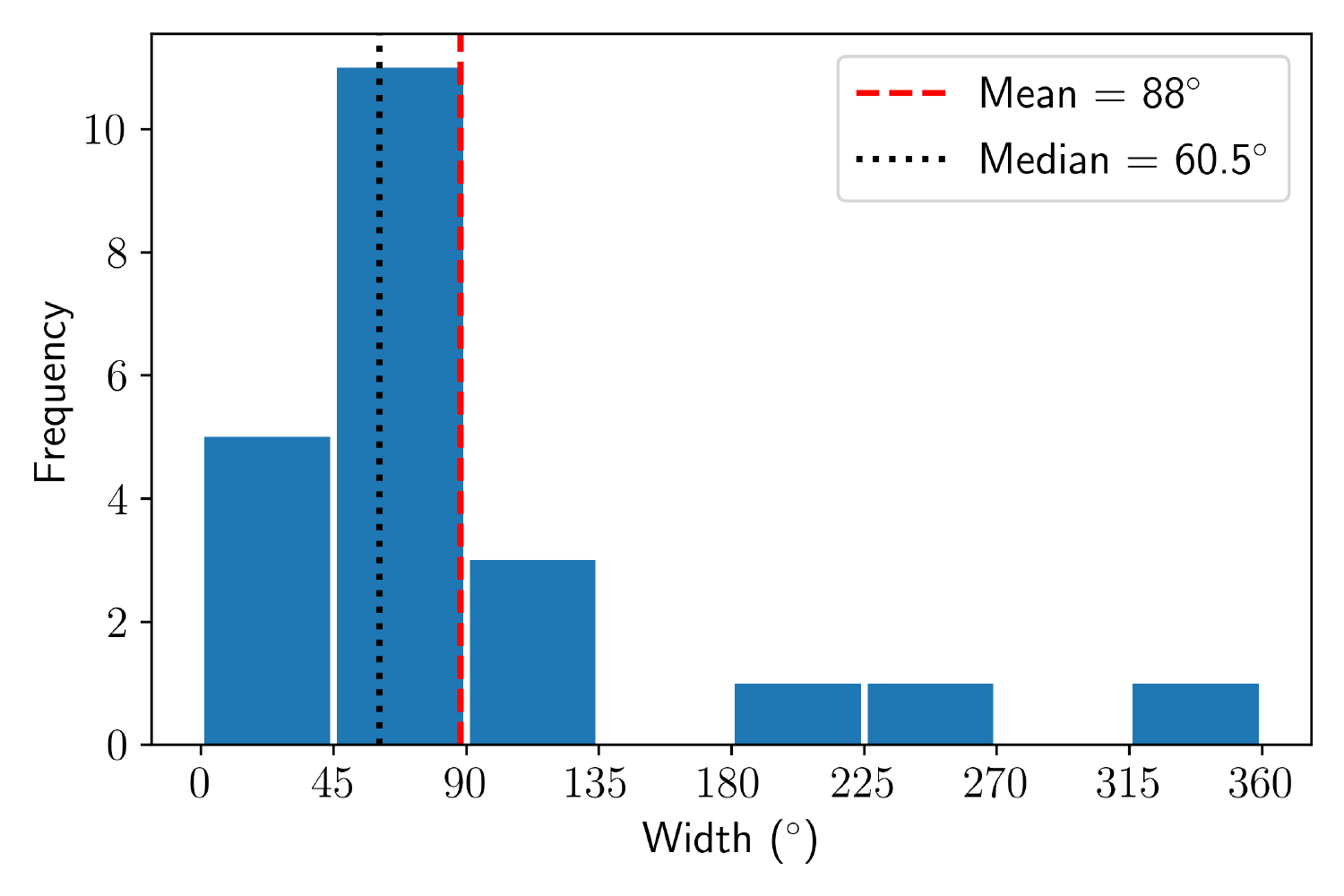}
\caption{Histogram of the PAs of the streamer axis for all streamer wave events reported in Table~\ref{tab:events} in the upper left panel. The upper right pannel shows the bar plot of all the measured shifts of the PA of the streamer axis between before and after the event. Histogram of the absolute value of the difference between the PA of the streamer axis and the CPA of the CME is shown in the lower left panel. The lower right panel shows the histogram of the widths of the CMEs reported in Table~\ref{tab:events}.}
\label{fig:group1}
\end{figure*}

We start our study \edit1{\replaced{with}{by}} analyzing the PA of the streamer axis for our streamer wave events. The upper left panel of Figure~\ref{fig:group1} shows a histogram of all the measured PAs for our streamer wave events and the colors indicate how they are distributed into equatorial and polar streamers. A streamer is called equatorial when the PA of the streamer axis is located less than 45\degr{} from the solar equatorial plane. From this histogram we can see that there is no preferred location for the streamers found in our survey. The streamer PAs vary from completely northern (0\degr{} for event 12), to completely equatorial towards the east (90\degr{} for event 13), to almost completely southern (175\degr{} for event 18), and to almost completely equatorial to the west (277\degr{} for event 3). There are almost an equal number of equatorial and polar streamers, as can be seen from the color of the bars in the upper left panel of Figure~\ref{fig:group1}. 

The time-distance map shown in Figure~\ref{fig:timedistance} shows that for the event on 2013-02-06 the position of the streamer axis after the wave event does not seem to match the position before the event. There is a shift of 2\degr{} to the south in the PA. In 17 cases the streamer axis has shifted from its original position as can be seen in the upper right panel of Figure~\ref{fig:group1}. The shift ranges from 1\degr{} to 5\degr{} and on average the streamer axis is displaced by 1.64\degr{} (this takes into account the 0\degr{} shift for 5 events). The displacement of the streamer axis is always directed away from the origin of the CME that perturbs the streamer. Therefore we believe that this shift is directly related to the CME shock displacing the streamer as a whole and to the large-scale restructuring of the corona during the CME.

To look at the positional relationship of the CME and the streamer, we compare the measurements of the PA of the streamer axis and the CPA of the CME in the FOV of the main coronagraph. We take the absolute value of the difference between these two values. If this difference is larger than 180\degr{}, then we subtract it from 360\degr{} to obtain the smaller value. In the lower left panel of Figure~\ref{fig:group1} we present the histogram of these differences between the two position angle measurements. The minimum difference between the PA of the streamer axis and the CPA of the CME is 18\degr{} and the maximum is 155\degr{}. The average difference is 58.8\degr{} with a standard deviation of 31.4\degr{}. We see that most of the streamer wave events occur when the streamer axis and the CME CPA are closer than 90\degr{} in the FOV of the coronagraph. Only three events have a difference in PA of more than 90\degr{}. This indicates that the relative position of the streamer and the CME origin may have an influence on whether or not a streamer event will occur, even if the PA in the coronagraph FOV does not directly show the true distance between the streamer and the CME origin.  

The apparent angular widths of the CMEs in the plane of the sky are shown in a histogram in the lower right panel of Figure~\ref{fig:group1}. The narrowest CME is 21\degr{} wide and the maximum width is 360\degr{} (full halo CME). The average width of our CMEs is 88\degr{} with a standard deviation of 21\degr{}, the median is 60.5\degr{}. This is much lower than the value for the apparent angular width of 290\degr{} reported by \citet{feng_streamer_2011}. The histogram of our events in the lower right panel of Figure~\ref{fig:group1} shows that most of our events are narrower than 120\degr{}. This shows that CMEs of all widths can excite a streamer wave event.

\begin{figure*}
\centering
\includegraphics[width=\textwidth]{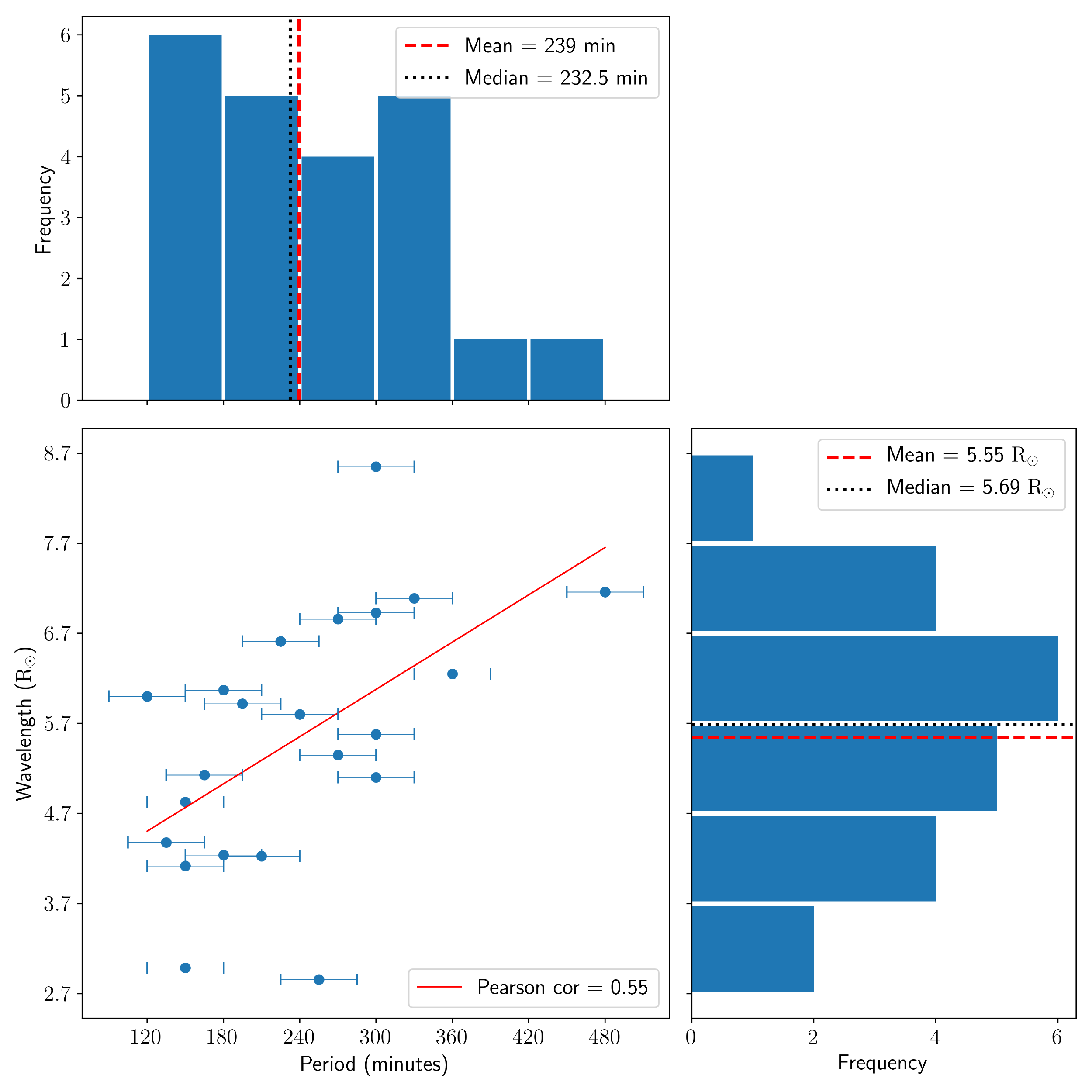}
\caption{Histogram of the periods (upper left) and wavelengths (lower right) for all streamer wave events reported in Table~\ref{tab:events}. Bottom left panel: scatterplot of the period versus the wavelength for all streamer wave events reported in Table~\ref{tab:events}. The linear regression line fit to the data is shown in red.}
\label{fig:group2}
\end{figure*}

The periods of the streamer wave events range from 2 hours up to 8 hours (see the upper left panel of Figure~\ref{fig:group2}). When discussing the periods, we have to take into account that the measurement errors can go up to half an hour due to the cadence of the COR2 coronagraphs that is 15 minutes. The average period of all streamer wave events is 239 minutes, with a standard deviation of 88 minutes. The median period is 232.5 minutes. From the histogram in the upper left panel of Figure~\ref{fig:group2} we can see that most of the periods are quasi-uniformly distributed between 2 and 6 hours. Only two events have a period longer than 6 hours. 

The \edit1{\replaced{upper}{lower}} right panel of Figure~\ref{fig:group2} shows that the measured wavelengths vary between 2.86 $\mathrm{R}_{\odot}$ and 8.55 $\mathrm{R}_{\odot}$. From the histogram, we can see that the wavelengths have a Gaussian-like distribution. The mean of the wavelengths is 5.55 $\mathrm{R}_{\odot}$ with a standard deviation of 1.41 $\mathrm{R}_{\odot}$. The median is given by 5.69 $\mathrm{R}_{\odot}$. The range of the measured values for the wavelength makes these the largest periodic waves that have been observed in the solar corona to our knowledge.

We combine our measurements of the wavelength and the period for all events in a scatterplot given in the bottom \edit1{\added{left}} panel of Figure~\ref{fig:group2}. The Pearson correlation coefficient of the two quantities is 0.55, indicated by the red linear regression line fit in the plot. The period and the wavelength are thus moderately correlated, where we get generally larger wavelengths when the period of the wave becomes longer. This is an indication that the streamer wave is an eigenmode of the streamer plasma slab. \edit1{\added{For eigenmodes of a streamer slab, theoretical models can be developed since the wave then only depends on the properties of the streamer slab itself. Together with the observations presented here, the models would be very suitable for a coronal seismology study.}} We do have to note here that the period is technically not a continuous variable in our measurements due to the cadence of 15 minutes of COR2, while in reality the periods do vary between the distinct values that we have here.    

\begin{figure*}
\centering
\includegraphics[width=0.45\textwidth]{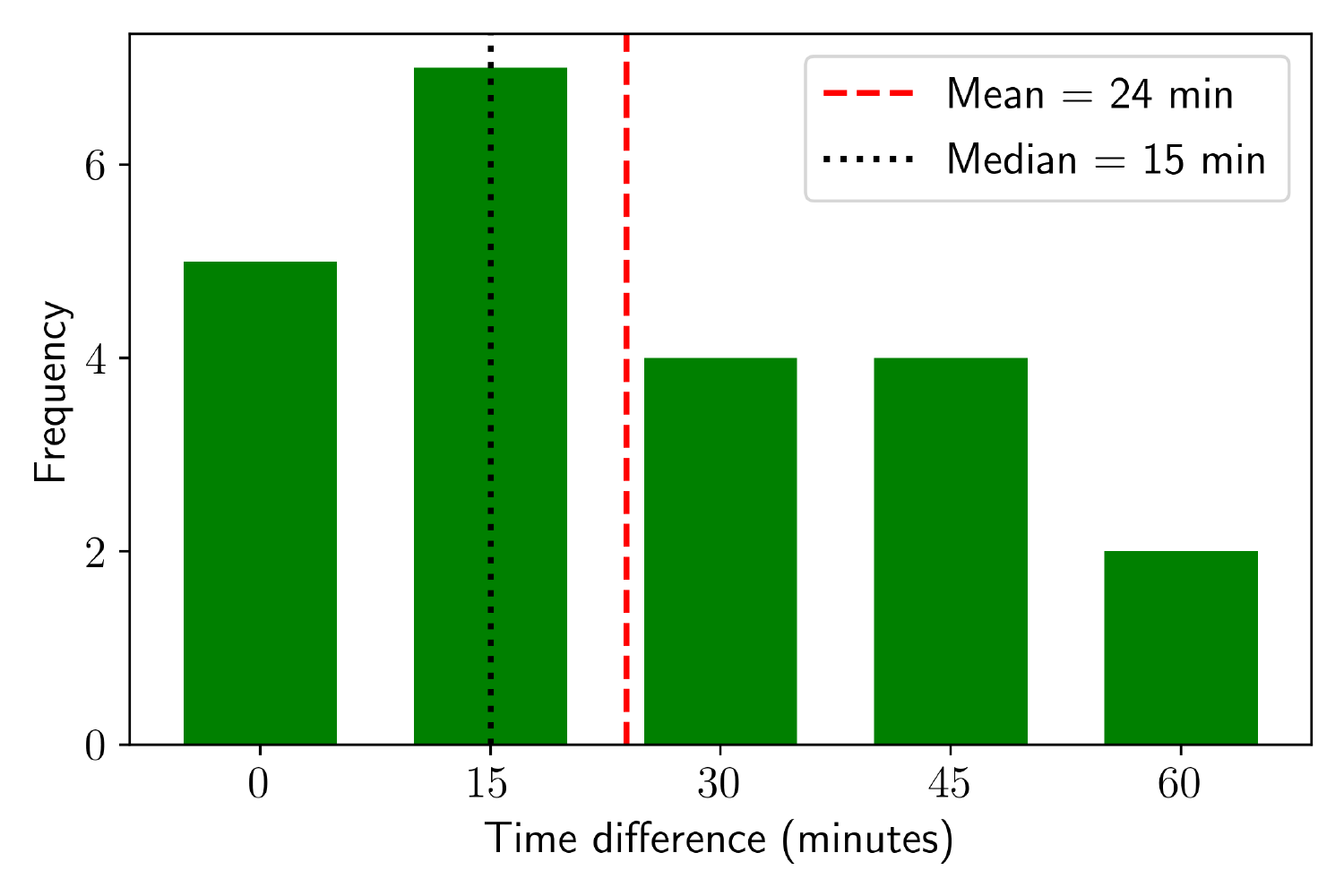}~
\includegraphics[width=0.45\textwidth]{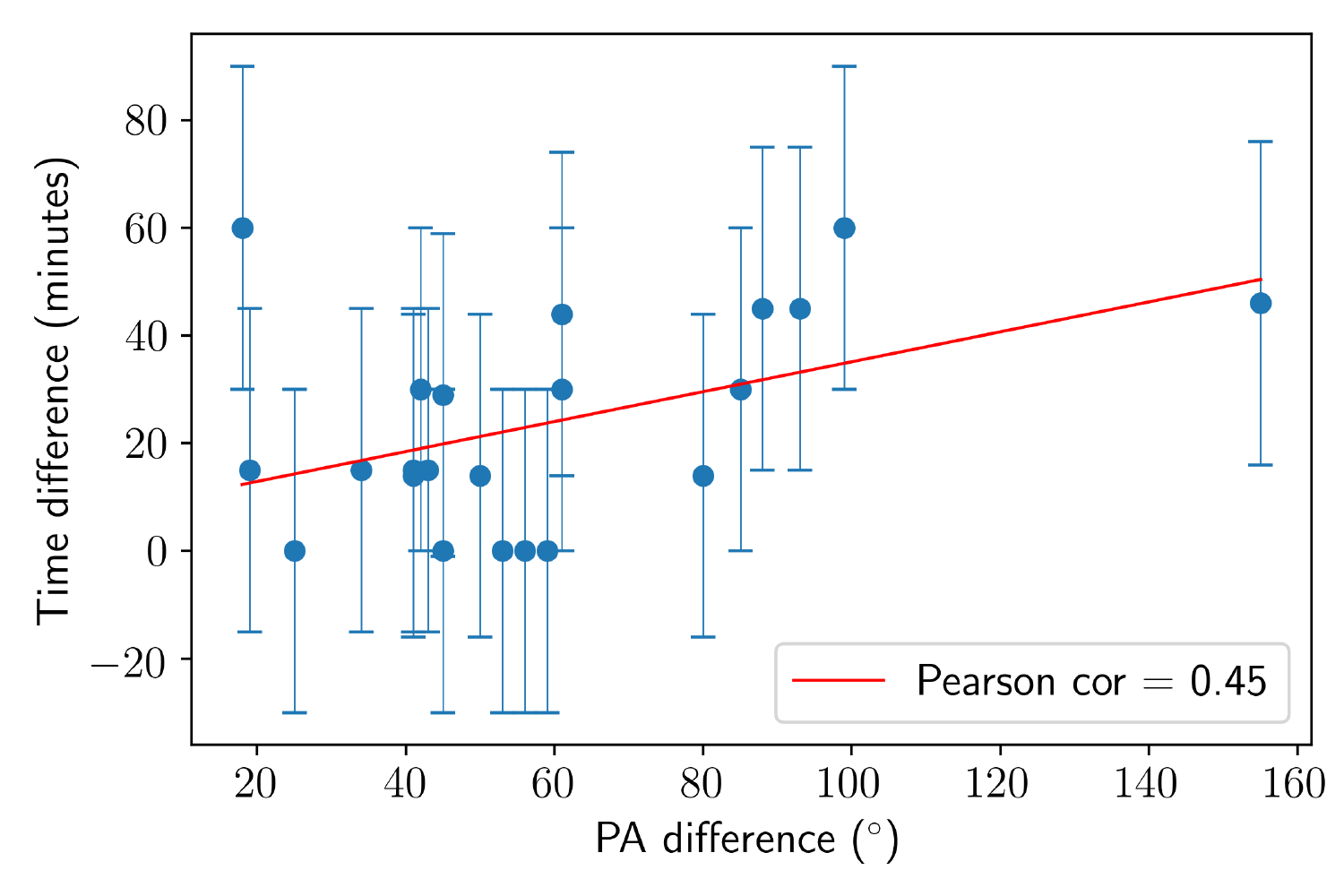}
\caption{Histogram of the difference in time between the first observation of the CME and the streamer wave (left) and scatterplot of the absolute value of the difference between the PA of the streamer axis and the CPA of the CME versus the difference in time between the first observation of the CME and the first observation of the streamer wave (right). The linear regression line fit to the data is shown in red.}
\label{fig:group3}
\end{figure*}

Next, we study the time difference between the first observation of the CME in the coronagraph FOV and the first observation of the streamer wave. Due to the cadence of the COR2 coronagraphs of 15 minutes, the delay between the two observations is divided into 5 categories from a minimum of 0 minutes, which means that the CME and the streamer wave are first visible in the same frame, to a maximum time difference between frames of 60 minutes (Figure~\ref{fig:group3}, left panel). The average time difference is 24 minutes with a standard deviation of 20 minutes. From the histogram in the left panel in Figure~\ref{fig:group3} we can see that a delay of 15 minutes between the two events is observed most often. For events in which the streamer wave and the CME are not first visible in the same time frame, the CME is always observed earlier than the streamer wave. When the streamer wave and the CME are first visible in the same frame, we can not tell for sure which of the two events occurred first. However, since we never observe the streamer wave event before the CME, we find it most likely that for all our events the streamer waves are caused by a perturbation of the streamer by the CME. 

The right panel of Figure~\ref{fig:group3} shows a scatter plot of the difference in the PA of the streamer axis and the CPA of the CME (as presented in the paragraph above) versus the difference in time between the first observation of the CME and the first observation of the streamer wave. The Pearson correlation coefficient between these two variables is 0.45, indicating a moderate correlation between them. When the source region of the CME that initiates the wave event is located farther from the streamer, it should take longer for the CME to reach the streamer and perturb it. This then results in a larger time difference between the first observation of the CME and the first observation of the streamer wave. Here, we only find a moderate correlation which could be due to the time uncertainty due to a low cadence of COR2 which is indicated by the error bars in the right panel of Figure~\ref{fig:group3}. Another possible cause is that the true distance between the CME source region and the streamer ``roots'' on the solar surface could be different from the distance inferred from the CME and streamer PAs in the plane of the sky. Finally, the expansion speed of each CME is different and could also play a role. If a CME expands faster, it will reach the streamer earlier.

\begin{figure*}
\centering
\includegraphics[width=\textwidth]{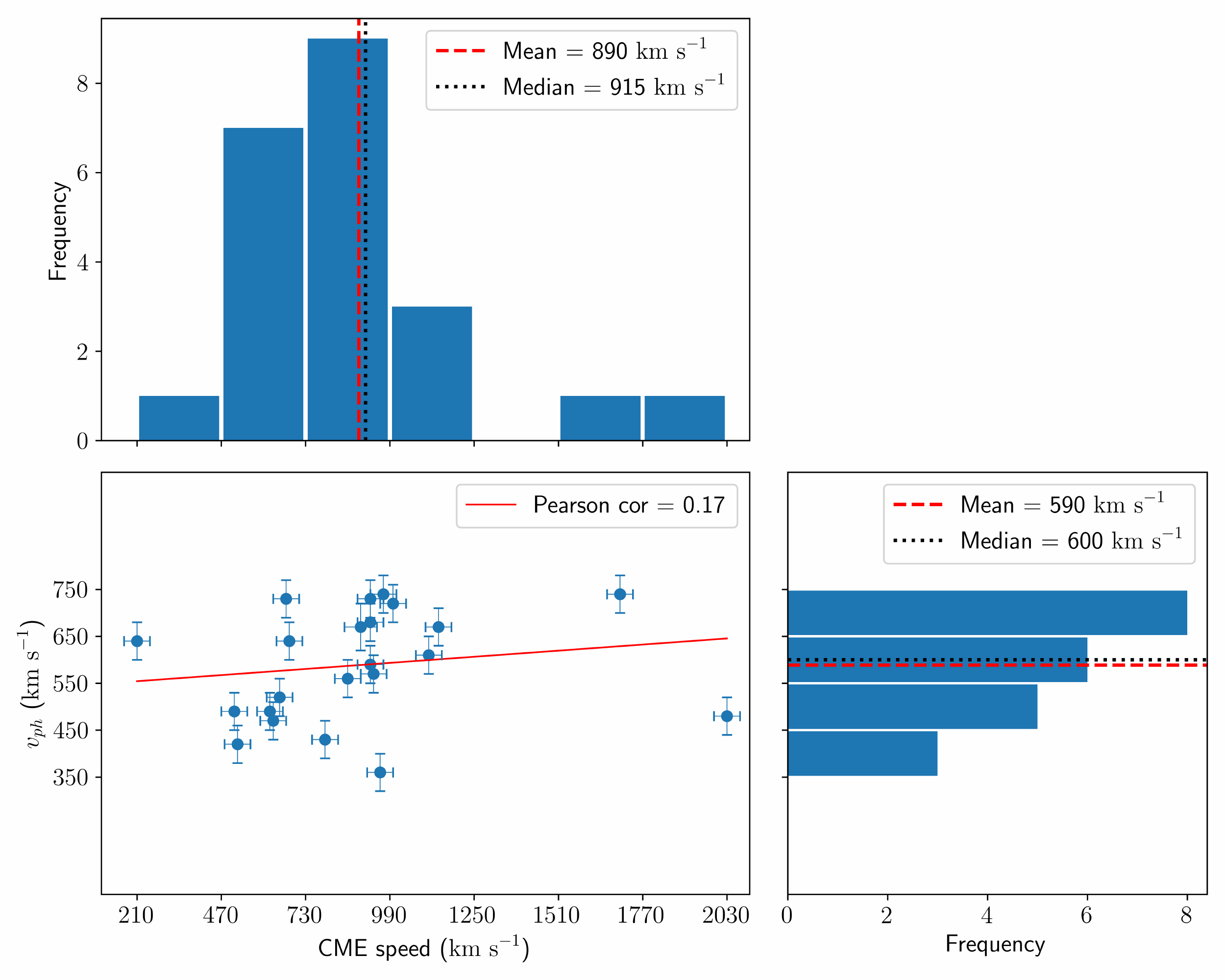}
\caption{Histogram of speeds of the CMEs (upper left) and the phase speeds of the streamer waves (lower right) reported in Table~\ref{tab:events}. The bottom left panel shows the scatterplot of the CME speed versus the phase speed of the corresponding streamer wave event. The linear regression fit to the data is shown in red.}
\label{fig:group4}
\end{figure*}

Another property of the streamer waves that we measured is the phase speed $v_{ph}$ of the waves (see the lower right panel of Figure~\ref{fig:group4}). The average of the phase speeds is 590 $\mathrm{km \ s}^{-1}$ with a standard deviation of 120 $\mathrm{km \ s}^{-1}$ and the median phase speed is 600 $\mathrm{km \ s}^{-1}$. There seems to be an increasing trend towards a speed around 700 $\mathrm{km \ s}^{-1}$, but the speeds present a cut-off around 750 $\mathrm{km \ s}^{-1}$. The slowest wave that we found has a phase speed of 360 $\mathrm{km \ s}^{-1}$ and the fastest wave has a phase speed of 740 $\mathrm{km \ s}^{-1}$. The phase speeds thus all lie in a rather narrow range of only 380 $\mathrm{km \ s}^{-1}$ wide. The narrow range of the phase speeds indicate that they all seem to behave as a typical eigenmode of the streamer, which makes these waves good candidates for coronal seismology.  

The linear speeds of the CMEs range between 210 $\mathrm{km \ s}^{-1}$ and 2030 $\mathrm{km \ s}^{-1}$ (see the upper left panel of Figure~\ref{fig:group4}), with an average speed of 890 $\mathrm{km \ s}^{-1}$ and a standard deviation of 387 $\mathrm{km \ s}^{-1}$. The median speed is 915 $\mathrm{km \ s}^{-1}$. The CME speed range is much wider than the range of the phase speed of the streamer waves. The bottom left panel of Figure~\ref{fig:group4} shows a scatter plot of the wave phase speed and the CME speed. The Pearson correlation coefficient of 0.17 indicates a very poor correlation between the two properties. This indicates that the streamer wave speed is determined by the physical properties of the streamer rather than the properties of the CME. This result indicates that the streamer waves are excellent candidates for coronal seismology.

Event 8 stands out here as it is the only event where the CME that we identified as the initiator of the wave has a significantly lower speed than the phase speed of the streamer wave (see Table~\ref{tab:events}). This seems contrary to earlier results reported by \citet{feng_streamer_2011}, who found that all the initiating CMEs had very high speeds ($>$ 1000 $\mathrm{km \ s}^{-1}$). Event 8 is however the only event with a speed below 500 $\mathrm{km \ s}^{-1}$. In our study, the CMEs tend to be fast, with more than half of them having a speed of more than 860 $\mathrm{km \ s}^{-1}$ (see Table~\ref{tab:events} and Figure~\ref{fig:group4}, top left). We do however also find that slower CMEs can initiate a streamer wave event. Taking into account the result about the apparent angular width of the CMEs discussed above, we find that, in general, we can not identify any necessary condition for a CME to excite a streamer wave event. We believe that whether or not a streamer wave is excited by a CME depends more on the 3D structuring of the solar corona and in particular of the streamer.

\begin{figure}
\centering
\includegraphics[width=0.45\textwidth]{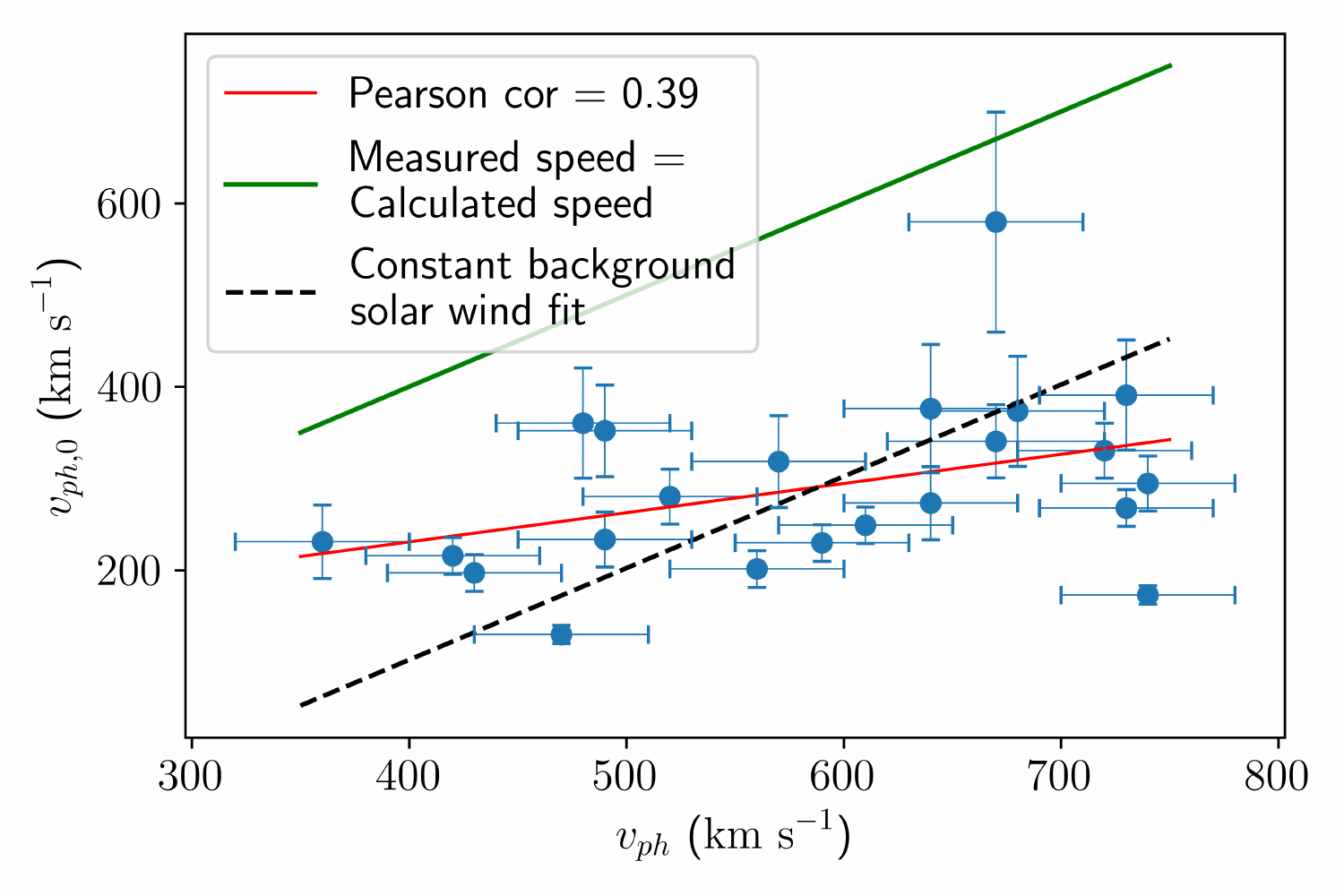}
\caption{Scatterplot of the phase speed calculated from the measured period and wavelength versus the measured phase speed. The linear regression fit to the data is shown in red and the green line corresponds to equal values of the calculated and the measured phase speed. The black line corresponds to a fit with a constant solar background wind of 298 $\mathrm{km \ s}^{-1}$.}
\label{fig:calcvsmeas}
\end{figure}

Finally, we can calculate the phase speed $v_{ph, 0}$ from the measured periods ($T$) and wavelengths ($\lambda$) through the relationship $v_{ph, 0} = \lambda / T$. In Figure~\ref{fig:calcvsmeas} the phase speeds are calculated for our events and compared to the measured phase speeds ($v_{ph}$) in a scatterplot. We can see that the measured speeds are consistently significantly higher than the calculated speeds. This is because our streamer waves are propagating in an already moving medium, namely the background solar wind in the streamer, which increases our measured speeds \citep[this increase is sometimes called a Doppler shift in MHD wave theory, see e.g. ][]{goossens_1992, nakariakov_1996}. \citet{chen_streamer_2012} and \citet{feng_streamer_2011} took this into account by subtracting a background solar wind profile from their measured phase speeds to obtain the phase speed of the wave in the plasma rest frame. If we assume that the background solar wind $v_{sw}$ is the same for all our streamer wave events, we can derive the average value for this solar wind speed. The best fit to our data of the form $v_{ph, 0} = v_{ph} - v_{sw}$ with a constant $v_{sw}$ is given by the black dashed line in Figure~\ref{fig:calcvsmeas}. This fit gives us a background solar wind speed of 298 $\mathrm{km \ s}^{-1}$, which is also the average of the difference between the measured and the calculated phase speeds. The minimum difference between the two speeds is 90 $\mathrm{km \ s}^{-1}$ and the maximum difference is 567 $\mathrm{km \ s}^{-1}$. From the linear regression fit to the scatter plot itself (red line in Figure~\ref{fig:calcvsmeas}) we can see that there is a moderate correlation with a Pearson correlation coefficient of 0.39. This indicates that the background solar wind speed will not be the same in all the streamers, but will also not vary too widely. 

The value obtained here for the solar wind speed in streamers indicates that streamers are, as expected, a source of the slow solar wind \edit1{(see e.g. \citealp{hundhausen_1977,gosling_1981,mccomas_1998}; and the recent review by \citealp{abbo_2016})}. However, the speed around 300 $\mathrm{km \ s}^{-1}$ that we found is higher than the slow wind speeds reported on the \edit1{\replaced{base}{basis}} of measurements of blobs by \edit1{\citet{sheeley_blobs_1997,wang_blobs_2000,sheeley_inout_2007}}\edit1{\added{; and \citet{jones_2009}}}. This is probably due to the transient increase of the solar wind speed in streamers after the passage of the CME-driven shock.

\section{Discussion and conclusions}\label{s:concl}
We made a survey of all streamer wave events observed by the COR2 coronagraphs aboard \textit{STEREO A} and \textit{B} between January 2007 and September 2017. This includes the duration of the \textit{STEREO} mission, when the white-light coronagraphic observations from three (two after 2014) vantage points could be analyzed (including \textit{SOHO}/LASCO \edit1{\added{C2+C3}}). In total, we found 22 events during this period. For all streamer wave events identified in this study, we could find a CME that most probably produces the wave. There seem to be more events during the solar cycle maximum, namely in 2011, 2012, and 2013. Some events are visible from two or three viewpoints, but not all. According to our analysis, this is caused by the streamer wave events only being visible from a specific viewpoint. It is possible that a streamer wave is \edit1{\replaced{better}{best}} visible if a corresponding streamer slab is well aligned with the line of sight (head-on view) and thus the oscillation is situated in the plane of the sky.

For each streamer wave event we measured the position angle, period, wavelength, and phase speed from the observations. Also, we identified the CME that excited the streamer wave and measured its central position angle, angular width, and linear plane-of-the-sky radial speed in the FOV of the COR2 coronagraph that observed the streamer wave event best.

We performed statistical analysis of the data set of measured properties of the streamer wave events and corresponding CMEs. We found that there are no preferred locations for the axis of the oscillating streamer. The streamer waves have long periods and very large wavelengths, making them one of the largest periodic wave phenomena observed in the solar corona. Streamer waves with larger wavelengths tend to have longer periods. The phase speeds of the streamer waves belong to a rather narrow range between 360 and 740 $\mathrm{km \ s}^{-1}$. We could not identify any necessary condition under which a CME is to excite a streamer wave. The CMEs in our survey do tend to be fast, but we also found slow CMEs that initiate a streamer wave event. The most important conclusion is that there appears to be no relation between the speed of the CME and the phase speed of the resulting streamer wave. This implies that the wave properties are determined by the physical properties of the plasma and the magnetic field in the coronal streamer rather than by the properties of the incident CME. By comparing our measured phase speeds with phase speeds calculated from the measured periods and wavelengths, we also could derive that the average speed of the post\edit1{\replaced{)}{-}}shock solar wind in our streamers is around 300 $\mathrm{km \ s}^{-1}$.

The good correlation between the measured wavelengths and the periods indicates that the streamer waves are an eigenmode of the streamer plasma slabs. Together with the narrow range of the measured phase speeds and the lack of a relation between the phase speeds and the CME speeds, this suggests that streamer wave events are good candidates for coronal seismology techniques. They thus are suitable to determine plasma properties inside coronal streamers through connecting them with models for wave propagation. By modeling the streamer wave events and combining the models with the measurements obtained in this study, one could extract some physical parameters that are hard to extract directly from observations, like the solar wind speed and the magnetic field strength. First attempts at this have already been undertaken by \citet{chen_streamer_2012} and \citet{feng_streamer_2011} who used fairly simple models to describe streamer wave events observed by \textit{SOHO}/LASCO \edit1{\added{C2+C3}}. Another example of coronal seismology is the determination of the average solar wind speed in streamers reported in Section~\ref{s:stats}. A next step would be to develop specific and more adequate models for the streamer waves and connect these with the observations presented here. This could give a better insight into the physical properties of the plasma and the magnetic field in coronal streamers. \edit1{\added{The theoretical models provide a relation between the different physical parameters of the streamer slab. From the observations and measurements presented here, we know several of these parameters like the wavelength and phase speed. Estimates for the density in the streamer slab can be obtained from the density models that are available for fitting to white-light images \citep[see e.g.][]{decraemer_2019}. Then the theoretical models can be used to obtain an estimate for the still unknown parameters, such as the magnetic field strength and the background solar wind speed.}}

To confirm that it is really the distance between the origin of the CME and the streamer location that determines the starting time of the streamer wave event, one would need to examine the 3D location of the streamer with respect to the CME source region. 3D reconstructions of the large-scale coronal structure in our events should be plausible since all events were found in the course of the \textit{STEREO} mission. One would also need to connect the CME and streamers observed in white-light coronagraph images to underlying structures in extreme ultraviolet (EUV) images which give a view of the low corona closer to the solar surface. It is however difficult to make this connection with the current instrumentation. The FOVs of externally occulted coronagraphs (like COR2) and EUV imagers (like the Extreme UltraViolet Imager aboard \textit{STEREO}) have a significant gap between them that is covered only by internally occulted coronagraphs (like \textit{STEREO}/COR1), which are prone to high straylight and do not allow observations of fine coronal structures at sufficient resolution. Since streamers and CMEs are very bright structures, COR1 could provide additional information only about the large-scale structuring of the corona. Future experiments like the ASPIICS (Association of Spacecraft for Polarimetric and Imaging Investigation of the Corona of the Sun) coronagraph aboard the \textit{PROBA-3} (\textit{Project for On-Board Autonomy-3}) mission \citep{proba32, proba33, proba3} will fill this observational gap with information on the fine structures and have a potential to improve our knowledge of how the coronal streamers are connected to the typical cusp structures that one can observe in the low corona. 

The filling of this observational gap could also help to understand how the excitation of the streamer wave actually happens. For each of our events, the initial displacement of the streamer is observed to take place down to the inner limit of the FOV of the COR2 coronagraph. This means that the actual interaction between the streamer and the CME occurs below the COR2 occulter and can thus not be studied with the current coronagraph images only. Images of the lower corona, like EUV images or the images to be taken by ASPIICS, could aid in understanding this interaction if it is visible there.

%

\acknowledgments
This work was supported by a PhD grant awarded by the Royal Observatory of Belgium to one of the authors (BD). ANZ thanks the European Space Agency (ESA) and the Belgian Federal Science Policy Office (BELSPO) for their support in the framework of the PRODEX Programme. TVD was supported by the European Research Council (ERC) under the European Union's Horizon 2020 research and innovation programme (grant agreement No 724326) and the C1 grant TRACEspace of Internal Funds KU Leuven (number C14/19/089). The SECCHI data used here were produced by an international consortium of the Naval Research Laboratory (USA), Lockheed Martin Solar and Astrophysics Lab (USA), NASA Goddard Space Flight Center (USA), Rutherford Appleton Laboratory (UK), University of Birmingham (UK), Max-Planck-Institut for Solar System Research (Germany), Centre Spatiale de Li\`ege (Belgium), Institut d'Optique Th\'eorique et Appliqu\'ee (France), Institut d'Astrophysique Spatiale (France). The SOHO/LASCO data used here are produced by a consortium of the Naval Research Laboratory (USA), Max-Planck-Institut for Sonnensystemforschung (Germany), Laboratoire d'Astrophysique de Marseille (France), and the University of Birmingham (UK). SOHO is a project of international cooperation between ESA and NASA.

%
%
\bibliographystyle{aasjournal} 
\bibliography{references}

\begin{thebibliography}{}
\expandafter\ifx\csname natexlab\endcsname\relax\def\natexlab#1{#1}\fi
\providecommand{\url}[1]{\href{#1}{#1}}
\providecommand{\dodoi}[1]{doi:~\href{http://doi.org/#1}{\nolinkurl{#1}}}
\providecommand{\doeprint}[1]{\href{http://ascl.net/#1}{\nolinkurl{http://ascl.net/#1}}}
\providecommand{\doarXiv}[1]{\href{https://arxiv.org/abs/#1}{\nolinkurl{https://arxiv.org/abs/#1}}}

\bibitem[{{Abbo} {et~al.}(2016){Abbo}, {Ofman}, {Antiochos}, {Hansteen},
  {Harra}, {Ko}, {Lapenta}, {Li}, {Riley}, {Strachan}, {von Steiger}, \&
  {Wang}}]{abbo_2016}
{Abbo}, L., {Ofman}, L., {Antiochos}, S.~K., {et~al.} 2016, \ssr, 201, 55,
  \dodoi{10.1007/s11214-016-0264-1}

\bibitem[{{Andries} {et~al.}(2009){Andries}, {van Doorsselaere}, {Roberts},
  {Verth}, {Verwichte}, \& {Erd{\'e}lyi}}]{andries_2009}
{Andries}, J., {van Doorsselaere}, T., {Roberts}, B., {et~al.} 2009, \ssr, 149,
  3, \dodoi{10.1007/s11214-009-9561-2}

\bibitem[{{Arregui}(2015)}]{arregui_2015}
{Arregui}, I. 2015, Philosophical Transactions of the Royal Society of London
  Series A, 373, 20140261, \dodoi{10.1098/rsta.2014.0261}

\bibitem[{{Bemporad} {et~al.}(2008){Bemporad}, {Poletto}, {Landini}, \&
  {Romoli}}]{bemporad_2008}
{Bemporad}, A., {Poletto}, G., {Landini}, F., \& {Romoli}, M. 2008, Annales
  Geophysicae, 26, 3017, \dodoi{10.5194/angeo-26-3017-2008}

\bibitem[{{Bemporad} {et~al.}(2010){Bemporad}, {Soenen}, {Jacobs}, {Landini},
  \& {Poedts}}]{bemporad_2010}
{Bemporad}, A., {Soenen}, A., {Jacobs}, C., {Landini}, F., \& {Poedts}, S.
  2010, \apj, 718, 251, \dodoi{10.1088/0004-637X/718/1/251}

\bibitem[{{Brueckner} {et~al.}(1995){Brueckner}, {Howard}, {Koomen},
  {Korendyke}, {Michels}, {Moses}, {Socker}, {Dere}, {Lamy}, {Llebaria},
  {Bout}, {Schwenn}, {Simnett}, {Bedford}, \& {Eyles}}]{brueckner_large_1995}
{Brueckner}, G.~E., {Howard}, R.~A., {Koomen}, M.~J., {et~al.} 1995, \solphys,
  162, 357, \dodoi{10.1007/BF00733434}

\bibitem[{{Chen} {et~al.}(2011){Chen}, {Feng}, {Li}, {Song}, {Xia}, {Kong}, \&
  {Li}}]{chen_streamer_2012}
{Chen}, Y., {Feng}, S.~W., {Li}, B., {et~al.} 2011, \apj, 728, 147,
  \dodoi{10.1088/0004-637X/728/2/147}

\bibitem[{{Chen} {et~al.}(2010){Chen}, {Song}, {Li}, {Xia}, {Wu}, {Fu}, \&
  {Li}}]{chen_streamer_2010}
{Chen}, Y., {Song}, H.~Q., {Li}, B., {et~al.} 2010, \apj, 714, 644,
  \dodoi{10.1088/0004-637X/714/1/644}

\bibitem[{{Cho} {et~al.}(2008){Cho}, {Bong}, {Kim}, {Moon}, {Dryer},
  {Shanmugaraju}, {Lee}, \& {Park}}]{cho_2008}
{Cho}, K.~S., {Bong}, S.~C., {Kim}, Y.~H., {et~al.} 2008, \aap, 491, 873,
  \dodoi{10.1051/0004-6361:20079013}

\bibitem[{{Cho} {et~al.}(2007){Cho}, {Lee}, {Gary}, {Moon}, \&
  {Park}}]{cho_2007}
{Cho}, K.~S., {Lee}, J., {Gary}, D.~E., {Moon}, Y.~J., \& {Park}, Y.~D. 2007,
  \apj, 665, 799, \dodoi{10.1086/519160}

\bibitem[{{Cranmer}(2012)}]{cranmer_2012}
{Cranmer}, S.~R. 2012, \ssr, 172, 145, \dodoi{10.1007/s11214-010-9674-7}

\bibitem[{{De Moortel} \& {Nakariakov}(2012)}]{demoortel_2012}
{De Moortel}, I., \& {Nakariakov}, V.~M. 2012, Philosophical Transactions of
  the Royal Society of London Series A, 370, 3193,
  \dodoi{10.1098/rsta.2011.0640}

\bibitem[{{Decraemer} {et~al.}(2019){Decraemer}, {Zhukov}, \& {Van
  Doorsselaere}}]{decraemer_2019}
{Decraemer}, B., {Zhukov}, A.~N., \& {Van Doorsselaere}, T. 2019, \apj, 883,
  152, \dodoi{10.3847/1538-4357/ab3b58}

\bibitem[{{Dolla} {et~al.}(2012){Dolla}, {Marqu{\'e}}, {Seaton}, {Van
  Doorsselaere}, {Dominique}, {Berghmans}, {Cabanas}, {De Groof}, {Schmutz},
  {Verdini}, {West}, {Zender}, \& {Zhukov}}]{dolla_2012}
{Dolla}, L., {Marqu{\'e}}, C., {Seaton}, D.~B., {et~al.} 2012, \apjl, 749, L16,
  \dodoi{10.1088/2041-8205/749/1/L16}

\bibitem[{{Feng} {et~al.}(2013){Feng}, {Inhester}, \& {Gan}}]{feng_2013}
{Feng}, L., {Inhester}, B., \& {Gan}, W.~Q. 2013, \apj, 774, 141,
  \dodoi{10.1088/0004-637X/774/2/141}

\bibitem[{{Feng} {et~al.}(2012){Feng}, {Chen}, {Kong}, {Li}, {Song}, {Feng}, \&
  {Liu}}]{feng_2012}
{Feng}, S.~W., {Chen}, Y., {Kong}, X.~L., {et~al.} 2012, \apj, 753, 21,
  \dodoi{10.1088/0004-637X/753/1/21}

\bibitem[{{Feng} {et~al.}(2011){Feng}, {Chen}, {Li}, {Song}, {Kong}, {Xia}, \&
  {Feng}}]{feng_streamer_2011}
{Feng}, S.~W., {Chen}, Y., {Li}, B., {et~al.} 2011, \solphys, 272, 119,
  \dodoi{10.1007/s11207-011-9814-6}

\bibitem[{{Forbes}(2000)}]{forbes_2000}
{Forbes}, T.~G. 2000, \jgr, 105, 23153, \dodoi{10.1029/2000JA000005}

\bibitem[{{Galano} {et~al.}(2018){Galano}, {Bemporad}, {Buckley}, {Cernica},
  {D{\'a}niel}, {Denis}, {de Vos}, {Fineschi}, {Galy}, \& {Graczyk}}]{proba3}
{Galano}, D., {Bemporad}, A., {Buckley}, S., {et~al.} 2018, in Society of
  Photo-Optical Instrumentation Engineers (SPIE) Conference Series, Vol. 10698,
  Space Telescopes and Instrumentation 2018: Optical, Infrared, and Millimeter
  Wave, 106982Y, \dodoi{10.1117/12.2312493}

\bibitem[{{Goossens} {et~al.}(1992){Goossens}, {Hollweg}, \&
  {Sakurai}}]{goossens_1992}
{Goossens}, M., {Hollweg}, J.~V., \& {Sakurai}, T. 1992, \solphys, 138, 233,
  \dodoi{10.1007/BF00151914}

\bibitem[{{Gosling} {et~al.}(1981){Gosling}, {Borrini}, {Asbridge}, {Bame},
  {Feldman}, \& {Hansen}}]{gosling_1981}
{Gosling}, J.~T., {Borrini}, G., {Asbridge}, J.~R., {et~al.} 1981, \jgr, 86,
  5438, \dodoi{10.1029/JA086iA07p05438}

\bibitem[{{Howard} {et~al.}(2008){Howard}, {Moses}, {Vourlidas}, {Newmark},
  {Socker}, {Plunkett}, {Korendyke}, {Cook}, {Hurley}, {Davila}, {Thompson},
  {St Cyr}, {Mentzell}, {Mehalick}, {Lemen}, {Wuelser}, {Duncan}, {Tarbell},
  {Wolfson}, {Moore}, {Harrison}, {Waltham}, {Lang}, {Davis}, {Eyles},
  {Mapson-Menard}, {Simnett}, {Halain}, {Defise}, {Mazy}, {Rochus}, {Mercier},
  {Ravet}, {Delmotte}, {Auchere}, {Delaboudiniere}, {Bothmer}, {Deutsch},
  {Wang}, {Rich}, {Cooper}, {Stephens}, {Maahs}, {Baugh}, {McMullin}, \&
  {Carter}}]{howard_sun_2008}
{Howard}, R.~A., {Moses}, J.~D., {Vourlidas}, A., {et~al.} 2008, \ssr, 136, 67,
  \dodoi{10.1007/s11214-008-9341-4}

\bibitem[{{Hundhausen}(1977)}]{hundhausen_1977}
{Hundhausen}, A.~J. 1977, in Coronal Holes and High Speed Wind Streams, ed.
  J.~B. {Zirker}, 225--329

\bibitem[{{Hundhausen}(1993)}]{hundhausen_1993}
{Hundhausen}, A.~J. 1993, \jgr, 98, 13177, \dodoi{10.1029/93JA00157}

\bibitem[{{Jones} \& {Davila}(2009)}]{jones_2009}
{Jones}, S.~I., \& {Davila}, J.~M. 2009, \apj, 701, 1906,
  \dodoi{10.1088/0004-637X/701/2/1906}

\bibitem[{{Kaiser} {et~al.}(2008){Kaiser}, {Kucera}, {Davila}, {St. Cyr},
  {Guhathakurta}, \& {Christian}}]{kaiser_stereo_2008}
{Kaiser}, M.~L., {Kucera}, T.~A., {Davila}, J.~M., {et~al.} 2008, \ssr, 136, 5,
  \dodoi{10.1007/s11214-007-9277-0}

\bibitem[{{Kong} {et~al.}(2012){Kong}, {Chen}, {Li}, {Feng}, {Song}, {Guo}, \&
  {Jiao}}]{kong_2012}
{Kong}, X.~L., {Chen}, Y., {Li}, G., {et~al.} 2012, \apj, 750, 158,
  \dodoi{10.1088/0004-637X/750/2/158}

\bibitem[{{Koutchmy}(1971)}]{koutchmy_three_1971}
{Koutchmy}, S. 1971, \aap, 13, 79

\bibitem[{{Koutchmy} \& {Fagot}(1973)}]{koutchmy_1973}
{Koutchmy}, S., \& {Fagot}, J. 1973, \nat, 246, 414, \dodoi{10.1038/246414a0}

\bibitem[{{Koutchmy} \& {Livshits}(1992)}]{koutchmy_1992}
{Koutchmy}, S., \& {Livshits}, M. 1992, \ssr, 61, 393,
  \dodoi{10.1007/BF00222313}

\bibitem[{{Krishna Prasad} {et~al.}(2018){Krishna Prasad}, {Raes}, {Van
  Doorsselaere}, {Magyar}, \& {Jess}}]{prasad_2018}
{Krishna Prasad}, S., {Raes}, J.~O., {Van Doorsselaere}, T., {Magyar}, N., \&
  {Jess}, D.~B. 2018, \apj, 868, 149, \dodoi{10.3847/1538-4357/aae9f5}

\bibitem[{{Kwon} {et~al.}(2013){Kwon}, {Ofman}, {Olmedo}, {Kramar}, {Davila},
  {Thompson}, \& {Cho}}]{kwon_2013}
{Kwon}, R.-Y., {Ofman}, L., {Olmedo}, O., {et~al.} 2013, \apj, 766, 55,
  \dodoi{10.1088/0004-637X/766/1/55}

\bibitem[{{Lamy} {et~al.}(2010){Lamy}, {Dam{\'e}}, {Viv{\`e}s}, \&
  {Zhukov}}]{proba32}
{Lamy}, P., {Dam{\'e}}, L., {Viv{\`e}s}, S., \& {Zhukov}, A. 2010, in Society
  of Photo-Optical Instrumentation Engineers (SPIE) Conference Series, Vol.
  7731, \procspie, 773118, \dodoi{10.1117/12.858247}

\bibitem[{{Lamy} {et~al.}(2019){Lamy}, {Floyd}, {Miki{\'c}}, \&
  {Riley}}]{lamy_2019}
{Lamy}, P., {Floyd}, O., {Miki{\'c}}, Z., \& {Riley}, P. 2019, \solphys, 294,
  162, \dodoi{10.1007/s11207-019-1549-9}

\bibitem[{{Loucif} \& {Koutchmy}(1989)}]{loucif_solar_1989}
{Loucif}, M.~L., \& {Koutchmy}, S. 1989, \aaps, 77, 45

\bibitem[{{Magdaleni{\'c}} {et~al.}(2014){Magdaleni{\'c}}, {Marqu{\'e}},
  {Krupar}, {Mierla}, {Zhukov}, {Rodriguez}, {Maksimovi{\'c}}, \&
  {Cecconi}}]{magdalenic_2014}
{Magdaleni{\'c}}, J., {Marqu{\'e}}, C., {Krupar}, V., {et~al.} 2014, \apj, 791,
  115, \dodoi{10.1088/0004-637X/791/2/115}

\bibitem[{{McComas} {et~al.}(1998){McComas}, {Bame}, {Barraclough}, {Feldman},
  {Funsten}, {Gosling}, {Riley}, {Skoug}, {Balogh}, {Forsyth}, {Goldstein}, \&
  {Neugebauer}}]{mccomas_1998}
{McComas}, D.~J., {Bame}, S.~J., {Barraclough}, B.~L., {et~al.} 1998, \grl, 25,
  1, \dodoi{10.1029/97GL03444}

\bibitem[{{McLaughlin} {et~al.}(2018){McLaughlin}, {Nakariakov}, {Dominique},
  {Jel{\'\i}nek}, \& {Takasao}}]{mclaughlin_2018}
{McLaughlin}, J.~A., {Nakariakov}, V.~M., {Dominique}, M., {Jel{\'\i}nek}, P.,
  \& {Takasao}, S. 2018, \ssr, 214, 45, \dodoi{10.1007/s11214-018-0478-5}

\bibitem[{{Nakariakov} \& {Ofman}(2001)}]{nakariakov_2001}
{Nakariakov}, V.~M., \& {Ofman}, L. 2001, \aap, 372, L53,
  \dodoi{10.1051/0004-6361:20010607}

\bibitem[{{Nakariakov} {et~al.}(1999){Nakariakov}, {Ofman}, {Deluca},
  {Roberts}, \& {Davila}}]{nakariakov_1999}
{Nakariakov}, V.~M., {Ofman}, L., {Deluca}, E.~E., {Roberts}, B., \& {Davila},
  J.~M. 1999, Science, 285, 862, \dodoi{10.1126/science.285.5429.862}

\bibitem[{{Nakariakov} {et~al.}(1996){Nakariakov}, {Roberts}, \&
  {Mann}}]{nakariakov_1996}
{Nakariakov}, V.~M., {Roberts}, B., \& {Mann}, G. 1996, \aap, 311, 311

\bibitem[{{Nakariakov} \& {Verwichte}(2005)}]{nakariakov_2005}
{Nakariakov}, V.~M., \& {Verwichte}, E. 2005, Living Reviews in Solar Physics,
  2, 3, \dodoi{10.12942/lrsp-2005-3}

\bibitem[{{Newkirk}(1967)}]{newkirk_structure_1967}
{Newkirk}, Gordon, J. 1967, \araa, 5, 213,
  \dodoi{10.1146/annurev.aa.05.090167.001241}

\bibitem[{{Ofman}(2010)}]{ofman_2010}
{Ofman}, L. 2010, Living Reviews in Solar Physics, 7, 4,
  \dodoi{10.12942/lrsp-2010-4}

\bibitem[{{Pascoe} {et~al.}(2019){Pascoe}, {Hood}, \& {Van
  Doorsselaere}}]{pascoe_2019}
{Pascoe}, D.~J., {Hood}, A.~W., \& {Van Doorsselaere}, T. 2019, Frontiers in
  Astronomy and Space Sciences, 6, 22, \dodoi{10.3389/fspas.2019.00022}

\bibitem[{{Pneuman} \& {Kopp}(1971)}]{pneuman_1971}
{Pneuman}, G.~W., \& {Kopp}, R.~A. 1971, \solphys, 18, 258,
  \dodoi{10.1007/BF00145940}

\bibitem[{{Reiner} {et~al.}(2003){Reiner}, {Vourlidas}, {Cyr}, {Burkepile},
  {Howard}, {Kaiser}, {Prestage}, \& {Bougeret}}]{reiner_2003}
{Reiner}, M.~J., {Vourlidas}, A., {Cyr}, O.~C.~S., {et~al.} 2003, \apj, 590,
  533, \dodoi{10.1086/374917}

\bibitem[{{Renotte} {et~al.}(2015){Renotte}, {Alia}, {Bemporad}, {Bernier},
  {Bramanti}, {Buckley}, {Capobianco}, {Cernica}, {D{\'a}niel}, \&
  {Darakchiev}}]{proba33}
{Renotte}, E., {Alia}, A., {Bemporad}, A., {et~al.} 2015, in Society of
  Photo-Optical Instrumentation Engineers (SPIE) Conference Series, Vol. 9604,
  \procspie, 96040A, \dodoi{10.1117/12.2186962}

\bibitem[{{Saez} {et~al.}(2007){Saez}, {Llebaria}, {Lamy}, \&
  {Vibert}}]{saez_2007}
{Saez}, F., {Llebaria}, A., {Lamy}, P., \& {Vibert}, D. 2007, \aap, 473, 265,
  \dodoi{10.1051/0004-6361:20066777}

\bibitem[{{Saez} {et~al.}(2005){Saez}, {Zhukov}, {Lamy}, \&
  {Llebaria}}]{saez_2005}
{Saez}, F., {Zhukov}, A.~N., {Lamy}, P., \& {Llebaria}, A. 2005, \aap, 442,
  351, \dodoi{10.1051/0004-6361:20042016}

\bibitem[{{Sheeley} {et~al.}(2000){Sheeley}, {Hakala}, \&
  {Wang}}]{sheeley_shock_2000}
{Sheeley}, Jr., N.~R., {Hakala}, W.~N., \& {Wang}, Y.-M. 2000, \jgr, 105, 5081,
  \dodoi{10.1029/1999JA000338}

\bibitem[{{Sheeley} \& {Wang}(2007)}]{sheeley_inout_2007}
{Sheeley}, Jr., N.~R., \& {Wang}, Y.~M. 2007, \apj, 655, 1142,
  \dodoi{10.1086/510323}

\bibitem[{{Sheeley} {et~al.}(1997){Sheeley}, {Wang}, {Hawley}, {Brueckner},
  {Dere}, {Howard}, {Koomen}, {Korendyke}, {Michels}, {Paswaters}, {Socker},
  {St. Cyr}, {Wang}, {Lamy}, {Llebaria}, {Schwenn}, {Simnett}, {Plunkett}, \&
  {Biesecker}}]{sheeley_blobs_1997}
{Sheeley}, Jr., N.~R., {Wang}, Y.~M., {Hawley}, S.~H., {et~al.} 1997, \apj,
  484, 472, \dodoi{10.1086/304338}

\bibitem[{{Song} {et~al.}(2012){Song}, {Chen}, {Li}, {Kong}, \&
  {Feng}}]{song_2012}
{Song}, H.-Q., {Chen}, Y., {Li}, G., {Kong}, X.-L., \& {Feng}, S.-W. 2012,
  Physical Review X, 2, 021015, \dodoi{10.1103/PhysRevX.2.021015}

\bibitem[{{Van Doorsselaere} {et~al.}(2016){Van Doorsselaere}, {Kupriyanova},
  \& {Yuan}}]{vandoorsselaere_2016}
{Van Doorsselaere}, T., {Kupriyanova}, E.~G., \& {Yuan}, D. 2016, \solphys,
  291, 3143, \dodoi{10.1007/s11207-016-0977-z}

\bibitem[{{Wang} {et~al.}(2015){Wang}, {Ofman}, {Sun}, {Provornikova}, \&
  {Davila}}]{wang_2015}
{Wang}, T., {Ofman}, L., {Sun}, X., {Provornikova}, E., \& {Davila}, J.~M.
  2015, \apjl, 811, L13, \dodoi{10.1088/2041-8205/811/1/L13}

\bibitem[{{Wang} {et~al.}(2000{\natexlab{a}}){Wang}, {Sheeley}, \&
  {Rich}}]{wang_2000}
{Wang}, Y.~M., {Sheeley}, Jr., N.~R., \& {Rich}, N.~B. 2000{\natexlab{a}},
  \grl, 27, 149, \dodoi{10.1029/1999GL010698}

\bibitem[{{Wang} {et~al.}(2000{\natexlab{b}}){Wang}, {Sheeley}, {Socker},
  {Howard}, \& {Rich}}]{wang_blobs_2000}
{Wang}, Y.~M., {Sheeley}, Jr., N.~R., {Socker}, D.~G., {Howard}, R.~A., \&
  {Rich}, N.~B. 2000{\natexlab{b}}, \jgr, 105, 25133,
  \dodoi{10.1029/2000JA000149}

\bibitem[{{Wang} {et~al.}(1997){Wang}, {Sheeley}, {Howard}, {Kraemer}, {Rich},
  {Andrews}, {Brueckner}, {Dere}, {Koomen}, {Korendyke}, {Michels}, {Moses},
  {Paswaters}, {Socker}, {Wang}, {Lamy}, {Llebaria}, {Vibert}, {Schwenn}, \&
  {Simnett}}]{wang_1997}
{Wang}, Y.~M., {Sheeley}, Jr., N.~R., {Howard}, R.~A., {et~al.} 1997, \apj,
  485, 875, \dodoi{10.1086/304467}

\bibitem[{{Webb} \& {Howard}(2012)}]{webb_2012}
{Webb}, D.~F., \& {Howard}, T.~A. 2012, Living Reviews in Solar Physics, 9, 3,
  \dodoi{10.12942/lrsp-2012-3}

\bibitem[{{West} {et~al.}(2011){West}, {Zhukov}, {Dolla}, \&
  {Rodriguez}}]{west_2011}
{West}, M.~J., {Zhukov}, A.~N., {Dolla}, L., \& {Rodriguez}, L. 2011, \apj,
  730, 122, \dodoi{10.1088/0004-637X/730/2/122}

\bibitem[{{White} \& {Verwichte}(2012)}]{white_2012}
{White}, R.~S., \& {Verwichte}, E. 2012, \aap, 537, A49,
  \dodoi{10.1051/0004-6361/201118093}

\bibitem[{{Zhukov} {et~al.}(2008){Zhukov}, {Saez}, {Lamy}, {Llebaria}, \&
  {Stenborg}}]{zhukov_origin_2008}
{Zhukov}, A.~N., {Saez}, F., {Lamy}, P., {Llebaria}, A., \& {Stenborg}, G.
  2008, \apj, 680, 1532, \dodoi{10.1086/587924}

\end{thebibliography}

\listofchanges

\end{document}